\documentclass[rsi,graphicx, amsmath, amssymb, reprint]{revtex4-1}

\usepackage{xcolor}%
\usepackage{graphicx}%
\usepackage{xspace}%
\usepackage{wasysym}%
\usepackage{makecell}
\newcommand{\degree}{\ensuremath{^\circ}\xspace}

\newcommand{\hp}{\ensuremath{\mathrm{H}^+}\xspace}
\newcommand{\htp}{\ensuremath{\mathrm{H}_2^+}\xspace}
\newcommand{\hthreep}{\ensuremath{\mathrm{H}_3^+}\xspace}

\newcommand{\smco}{Sm$_2$Co$_{17}$\xspace}
\newcommand{\ndfe}{Nd$_2$Fe$_{14}$B\xspace}

\begin{document}


\title{The MIST-1 and MIST-2 multicusp ion sources\\ for high-current \htp beams} 




\author{D. Winklehner}%
\email[]{winklehn@mit.edu}
\author{S. Engebretson}
\author{J. Moon}
\author{J. Villarreal}
\author{P. Weigel}
\author{E. Winkler}
\affiliation{ 
Department of Physics, Massachusetts Institute of Technology, Cambridge, MA
}%
\author{P. Landon}
\affiliation{ 
Department of Physics, Boston University, Boston, MA
}%


\date{\today}

\begin{abstract}

We present two iterations of the \emph{Multicusp Ion Source Technology at MIT} (MIST) sources, designed to fulfill the requirements of the \emph{HCHC-XX} cyclotron design. The HCHC-XX is a novel compact cyclotron accelerating \htp. Beam is injected through a radio-frequency quadrupole buncher-accelerator (embedded in the cyclotron yoke) and utilizes so-called vortex motion during acceleration. If successful, it will deliver 10~mA of protons at 60~MeV in CW mode. This scheme requires
a low-emittance, high-current initial beam with high \htp purity.
We briefly summarize the design
and previous results of the MIST-1 ion source and, for the first time, the detailed design of 
the new and improved MIST-2, including the mechanical, electrical, and control system design. 
We further show experimental results of using the MIST-2 backplate on the MIST-1 body, present a study using different types of permanent magnets for confinement (including no magnets), and finally, we present first results of the MIST-2 in full operation. In this first commissioning run, we were able to increase the total extracted current from the MIST-2 to 7~mA--a factor of 2 over the MIST-1.
\end{abstract}


\maketitle 

\section{Introduction}
\label{sec:intro}
Our original incentive to develop multicusp ion sources lies in the proposed IsoDAR (Isotope Decay-At-Rest) experiment--a search for Beyond Standard Model (BSM) physics. IsoDAR will operate by placing an intense particle source
near a large underground scintillator detector~\cite{conradPrecisionAntineutrinoelectronScattering2014, alonsoNeutrinoPhysicsOpportunities2022,alonsoIsoDARYemilabConceptualDesign2021}. Arguably, one of the most exciting searches that IsoDAR can perform is for so-called \emph{sterile neutrinos}~\cite{winklehnerIsoDARYemilabADefinitiveSearch2023}.
If found, sterile neutrinos would resolve the long-standing discrepancy between predicted neutrino oscillation parameters and those observed in many experiments~\cite{conradSterileNeutrinos}. The IsoDAR experiment relies on a 10~mA continuous wave beam of 60~MeV/amu protons with an annual up-time of $>80$\%, impinging on a neutrino-production target, to obtain statistically significant results within five years. We plan to conduct this experiment in an underground environment to minimize the background from cosmic ray muons~\cite{bungauNeutrinoYieldNeutron2024}. We have developed the \emph{HCHC-XX} (High-Current \htp Cyclotron at XX MeV/amu) 
for IsoDAR~\cite{winklehnerIsoDARYemilabPreliminaryDesign2024}. Several novel features will permit the HCHC-XX family of cyclotrons to reach such high currents. These key unique features include:

\begin{itemize}
\setlength\itemsep{0.1em}
    \item Accelerating \htp
    \item Bunching through a Radio-Frequency Quadrupole (RFQ) axially embedded in the cyclotron~\cite{winklehnerHighCurrentH2Beams2021,holtermannTechnicalDesignRFQ2021}
    \item Taking advantage of vortex motion~\cite{stetson:vortex, baumgartenTransverselongitudinalCouplingSpace2011} during primary acceleration~\cite{winklehnerOrdermagnitudeBeamCurrent2022}
\end{itemize}

Following acceleration to 60 MeV/amu, we can split \htp into protons that can then be used in our physics experiments. Beyond particle physics, the HCHC-XX has several broader applications because of its high proton flux, including: applications to fusion material research~\cite{sneadEnablingMultiPurposeHighEnergy2023}, producing medical radioisotopes~\cite{winklehnerNewFamilyHighcurrent2024, alonsoMedicalIsotopeProduction2019}, and transmuting nuclear waste~\cite{ADS_Transmutation}.

Based on the envisioned compactness and the needs for the RFQ input beam and subsequent injection into the HCHC-XX through a so-called \emph{spiral inflector}~\cite{winklehnerRealisticSimulationsCyclotron2017}, we have identified the following ion source requirements:

\begin{itemize}
\setlength\itemsep{0.1em}
    \item 10~mA \htp current
    \item \htp ion fraction $\geq80\%$ 
    \item Norm. rms emittance  $\leq$ 0.1~$\pi$-mm-mrad
    \item Continuous (``DC'') beam
    \item Low energy spread
\end{itemize}

\begin{figure*}[!t]
\centering
\includegraphics[width=1.0\linewidth]{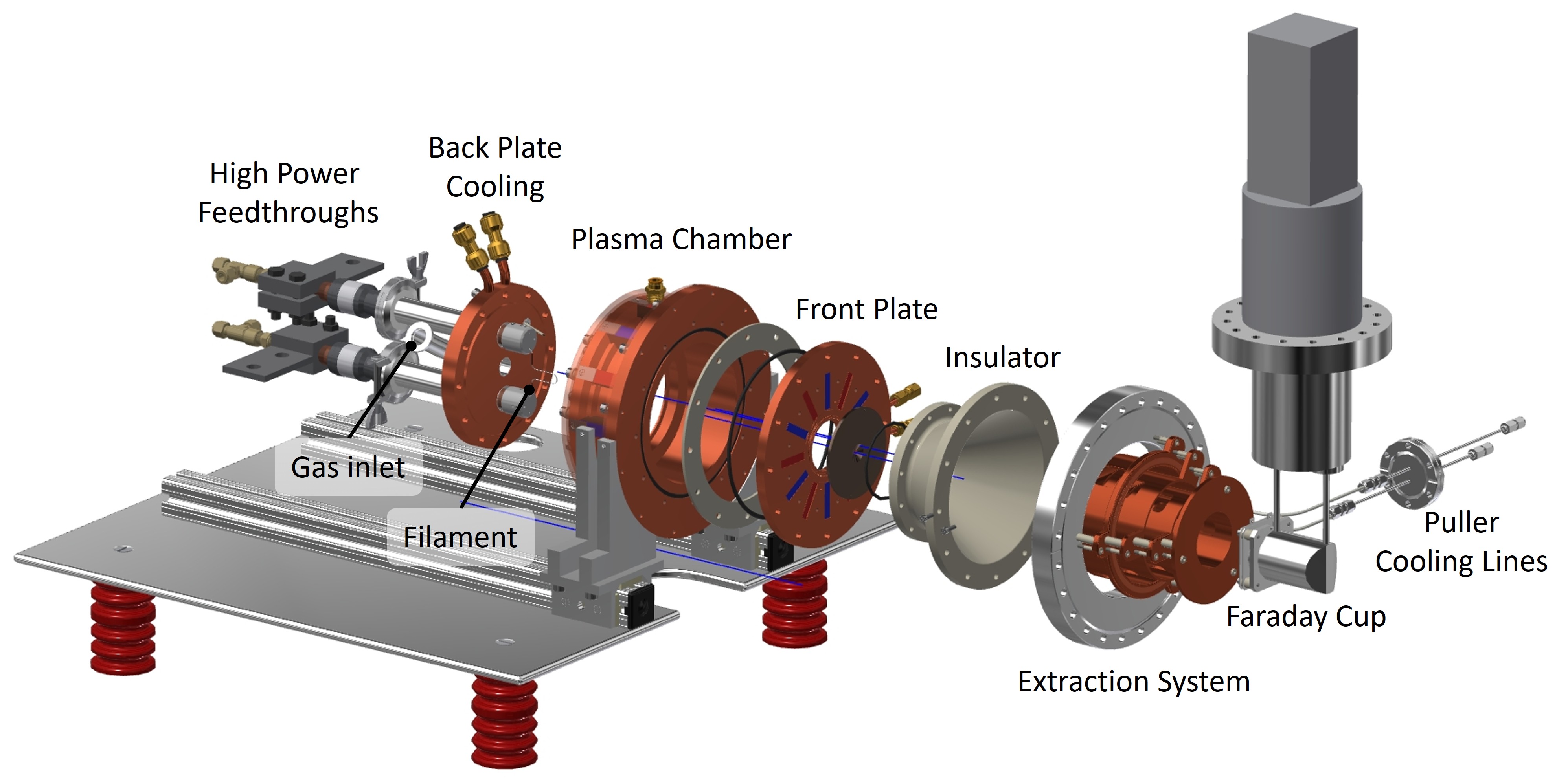}
\caption{MIST-2 CAD rendering--exploded view. Calling out the most important features. The water jacket around the plasma chamber is set to transparent in this view to show the radial confinement magnets. Radial and front plate magnets are indicated in red and blue (red - north, blue - south).}
\label{fig:MIST-2_Overview}
\end{figure*}

To satisfy all of them simultaneously, we settled on the filament-driven multicusp ion source type. 
In the present ion source landscape, there are a number of ion sources with exceptional performance in one or several of these metrics, but not, to our knowledge, all of them. For example, the miniaturized microwave ion source (MMIS) at Peking University produces a very large \htp current (16.1~mA) with $\sim50$\% \htp fraction~\cite{Peking_IonSource}. The MMIS is a 2.45~GHz electron cyclotron resonance (ECR) ion source with permanent magnet solenoids for confinement and providing the ECR resonance surface. Coincidentally, we have previously tested the Versatile Ion Source (VIS)~\cite{castroNewH2Source2016}, another, larger, 2.45~GHz ECR source built by INFN Catania, but we were not able to achieve an \htp fraction 
$>50\%$~\cite{alonsoCharacterizationCataniaVIS2014,alonsoIsoDARHighIntensity2015}. This finding is consistent with a general overview of the field presented in Ref.~\cite{wu:hydrogen}.
Another promising ion source is the filament-driven arc discharge ion source
for the FRANZ project~\cite{schweizerHighIntensity2002014} which has been able to produce a $91\%$ \htp fraction with 2.85~mA current~\cite{joshi:franz1}. This ion source had a relatively large energy spread of 120~eV. We chose to base our design on the multicusp ion source developed at LBNL, which had previously demonstrated $>80\%$ \htp ion fraction~\cite{ehlers:multicusp1}. Additionally, multicusp ion sources can have very low emittances as shown in Ref.~\cite{wutte:multicusp} which realized a normalized rms emittance of $<0.04~\pi$-mm-mrad for an argon beam. The energy spread can also be sufficiently low. See, for example, Ref.~\cite{lee_multicusp} reporting an energy spread below 2~eV for a hydrogen beam with a density of 12~$mA/cm^2$.

This paper proceeds as follows. In Section~\ref{sec:design} we present the design details of the MIST-2 ion source and highlight important differences to the previous version, MIST-1. We briefly present our
simulation pipeline in Section~\ref{sec:simulations}, followed by three experimental studies in Section~\ref{sec:measurement}. These studies are: Measurements with the new MIST-2 back plate on the original MIST-1 body; A study comparing different permanent magnet materials; The first commissioning results of the full MIST-2 ion source.

\section{The Design of the MIST ion sources}
\label{sec:design}

The design of the MIST-1 ion source was described in much detail in 
Refs.~\cite{winklehnerFirstCommissioningResults2018, winklehnerHighcurrentH2Beams2021a, winklehnerNewCommissioningResults2022}. MIST-2 was briefly described in the IsoDAR preliminary design report (PDR)~\cite{winklehnerIsoDARYemilabPreliminaryDesign2024}, but no details were given. Here, we will present in detail the mechanical and electrical design of MIST-2 and,
where relevant, the differences from MIST-1. For the full details of the MIST-1 design, we refer the reader to the references above.

In a multicusp ion source, the plasma electrons and ions are confined in a magnetic field produced by permanent magnet bars arranged in an alternating configuration 
(cf. Section~\ref{sec:source_body}). Ionization is facilitated by a hot filament emitting electrons, which are accelerated into the plasma by a potential applied between the filament and the source body. The dominant process for ionization is then electron impact ionization. Other atomic processes in the plasma are ionization by collision with other hadrons, dissociation by collision, and combination of 
ions to molecular ions. The balance of these competing processes determines the ion species fractions in the beam. Ions drifting towards the plasma aperture leave the ion source, as we intentionally keep the magnetic cusp field in this region low. At the plasma aperture, beam formation happens in an interplay between the plasma and the external electric field between source and puller (cf. Section~\ref{sec:simulations}). For a more detailed discussion on the
atomic processes in the plasma see Ref.~\cite{winklehnerHighcurrentH2Beams2021a} and references therein.

\subsection{Mechanical design}

\subsubsection{MIST-2 source body\label{sec:source_body}}
FIG.~\ref{fig:MIST-2_Overview} shows an exploded view of the MIST-2 ion source up to the
Faraday cup located 255~mm after the plasma aperture. The most important mechanical parameters
are listed in TABLE~\ref{tab:mist-2_params}. The plasma chamber, back plate, and front plate are made from oxygen-free high conductivity copper (OFHC), also known as Cu-101. The inside of the plasma chamber is 81~mm in length and 148~mm in diameter.
In addition to the short chamber, compared to its radius, we place the filament
close to the plasma aperture (the hole in the front of the source through which
ions drift out) at $\approx30$~mm.

The plasma chamber is surrounded by a \emph{water jacket}, a sleeve that seals a hollow space around the plasma chamber and that has a 3/8-inch push-to-connect water inlet on the bottom and a corresponding outlet on the top. The radial magnets
(see FIG.~\ref{fig:MIST-2_Overview}) sit in recesses around the chamber wall. There is a 2~mm gap underneath each magnet to let water pass between the magnets and the chamber wall. The radial magnets are grade N50H \ndfe permanent magnets of size  19~mm~$\times~13$~mm~$\times~50$~mm, placed with alternating polarity. The direction of magnetization is radially outward/inward.
Six magnets of the same size and quality are arranged in a checkerboard pattern on the back plate.
The magnets on the front plate are smaller at 8~mm~$\times~8$~mm~$\times~36$~mm and magnetized axially forward/backward. They are also placed with alternating polarity, as can be seen in FIG.~\ref{fig:MIST-2_Overview}. 
All magnets are nickel-coated for corrosion resistance. 

The center part of the front plate is made from a tungsten alloy (75\% W, 25\% Cu) that has good electrical conductivity as well as exceptional heat resistance. This plate can be replaced to test plasma apertures of different sizes.

The back plate, main chamber, and front plate are all individually cooled by de-ionized (DI) water and electrically insulated from each other using ceramic rings and Viton O-rings. We bake these O-rings before installation to reduce the beam contamination from outgassing of water and fluorine compounds into the plasma.

The source body is mounted to carriages that sit on two rails (1.5-inch aluminum extrusions) for easier installation and manipulation during component exchange.

The back plate features two KF NW-16 half-nipples. One to connect the mass flow controller (MFC), an MKS GV50A, with a full range of 5 SCCM (standard cubic centimeters per minute) of flow, calibrated for H$_2$, and RS-485 control. The other flange is a currently unused port for a Langmuir probe.
The mass flow controller is at ground potential and connected to the back plate via a 1/4-inch stainless steel tube and a 20~mm long ceramic break. The filament heating current is provided via to high-power (8~kV, 1~kA) water-cooled feedthroughs passing through KF NW-40 half-nipples. On the ends of these feedthroughs sit two molybdenum filament holders that can accept wires of different diameters. The filament is a pure tungsten wire. The current best-performing filament shape is a double-hairpin made out of 0.032" diameter wires. 

\begin{table}[b!]
\caption{The critical design parameters of the MIST-2 ion source.\label{tab:mist-2_params} }
\begin{tabular}{|l|l|}
\hline
\textbf{Parameter} & \textbf{Value} \\
\hline \hline
Body material & OFHC (Cu-101) \\
Plasma aperture material & 75\% W, 25\% Cu \\ 
Plasma aperture size & 6~mm \\
Plasma chamber size & 81~mm L $\times$ 148~mm \diameter \\
Cooling & Embedded DI Water channels \\
Total DI flow & $\approx10$~l/min \\
Permanent magnet material & \ndfe \\
Permanent magnet sizes & 18 pc. $19\times13\times50$~mm$^3$ \\
                       & 12 pc. $8\times8\times36$~mm$^3$ \\
Operating gas & H$_2$ \\
Gas control & MKS GV50A (5 SCCM) \\
Filament material & tungsten \\
Filament diameter & $\approx0.8$~mm \\
Filament shape & Double hairpin \\
\hline
\end{tabular}
\end{table}

\subsubsection{MIST-2 extraction system}
\begin{figure}[!t]
\centering
\includegraphics[width=1.0\columnwidth]{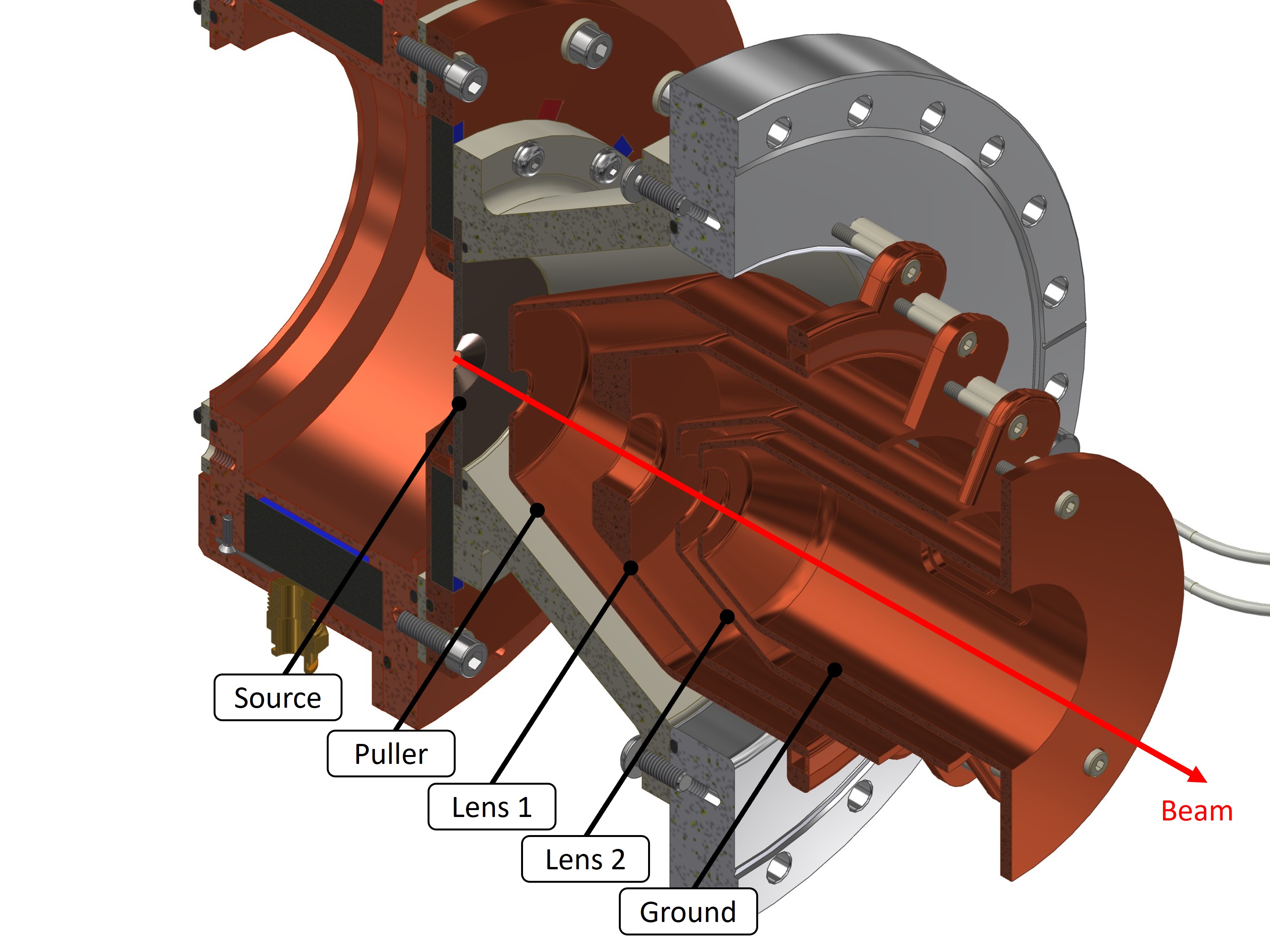}
\caption{MIST-2 Extraction system CAD rendering--cut view. The electrodes are labeled and the nominal voltages can be found in TABLE~\ref{tab:xs_nominal_voltages}.}
\label{fig:MIST-2_XS_CAD}
\end{figure}

We show a close-up cut view of the extraction system in FIG.~\ref{fig:MIST-2_XS_CAD} and list the nominal voltages for 10~mA total beam current extraction in TABLE~\ref{tab:xs_nominal_voltages}. These are based on the simulations presented in Section~\ref{sec:simulations} and correspond well to the values we found experimentally in Section~\ref{sec:measurement}. The source voltage is fixed at 15~kV. This is necessitated by the RFQ following later in the system, which requires an input beam energy of 15~keV. The puller is the first electrode after the source and
is set to a negative voltage to prevent electrons from back-streaming into the source. The puller can be varied from -1~kV to -10~kV to tune the electric field for extraction. Lens 1 and Lens 2 provide focusing, with Lens 2 being negative as well and providing the additional function of repelling electrons back into the low energy beam transport line (LEBT). The grounded final element shields the beam from the other high voltages and facilitates early space charge compensation (the gathering of low-energy electrons within the beam envelope, reducing space charge~\cite{winklehnerSpaceChargeCompensation2015}). 
The electrodes are made from OFHC copper and the extraction system is mounted on a CF~10-inch stainless steel flange. Alignment of the electrodes is facilitated through a kinematic mounting arrangement of three posts sitting inside precision-milled slots and three insulating screws per electrode. We align the extraction system with the main insulator and the source using stainless steel pins.
The puller is water-cooled using DI water going through a channel in its base.

\subsubsection{Design differences versus MIST-1}
The MIST-2 is an upgrade over the MIST-1, implementing lessons-learned from past runs.
The most significant differences in design are:
\begin{itemize}
\setlength\itemsep{0.1em}
    \item Change of main material for fabrication. The original choice of stainless steel
          for it's high yield strength and melting point proved problematic due to it being
          a bad heat conductor and making efficient cooling difficult.
    \item Cross-section increase in all cooling channels. In addition to the material, 
          we improved the amount of water that can be pumped through.
    \item New, more compact extraction system. This is not necessarily a core design change
          as we have tested a number of extraction systems. For MIST-2, we designed a new
          extraction system that facilitates space charge compensation early on and leaves 
          room in the first diagnostic 6-way cross to place a Faraday cup and, if desired,
          emittance scanners.
    \item New, higher power water-cooled feedthroughs for the filament. The MIST-1 feedthroughs
          provided adequate cooling, however, there were braze-joints to close the cooling loops inside the plasma chamber. We speculated that these may have caused some of
          the contaminants in in the beam. The new feedthroughs are off-the shelf, and are machined from a solid copper rod. They can easily be replaced and introduce no
          additional material into the source.
    \item Redesign of the front plate. The front plate was redesigned to move the cooling loop       closer to the plasma aperture, while at the same time placing the permanent magnets        entirely outside the vacuum. Thus a smoother surface is presented to the puller
          facilitating better high voltage performance.
\end{itemize}

\begin{table}[tbh!]
\caption{Nominal voltages of ion source platform and extraction system electrodes for MIST-2 at 10~mA total extracted current.\label{tab:xs_nominal_voltages} }
\begin{tabular}{|l|c|}
\hline
\textbf{Electrode} & \textbf{Voltage (kV)} \\
\hline \hline
Source & 15 \\
Puller & -8 \\
Lens 1 & 12 \\
Lens 2 & -3 \\
Ground & 0 \\
\hline
\end{tabular}
\end{table}

\subsection{Electrical design} 
The operation of the MIST-2 ion source and the associated test stand relies on a coordinated electrical system that provides power and control to all major subsystems. These include the source itself, the extraction assembly, the gas delivery system, and diagnostic systems. The power system must allow for precise, independent control of each subsystem while insulating components from each other and from the users. These elements form an integrated infrastructure enabling stable and efficient ion source operation.

\begin{figure}[!hbt]
  \centering
  \includegraphics[width=1.0\linewidth]{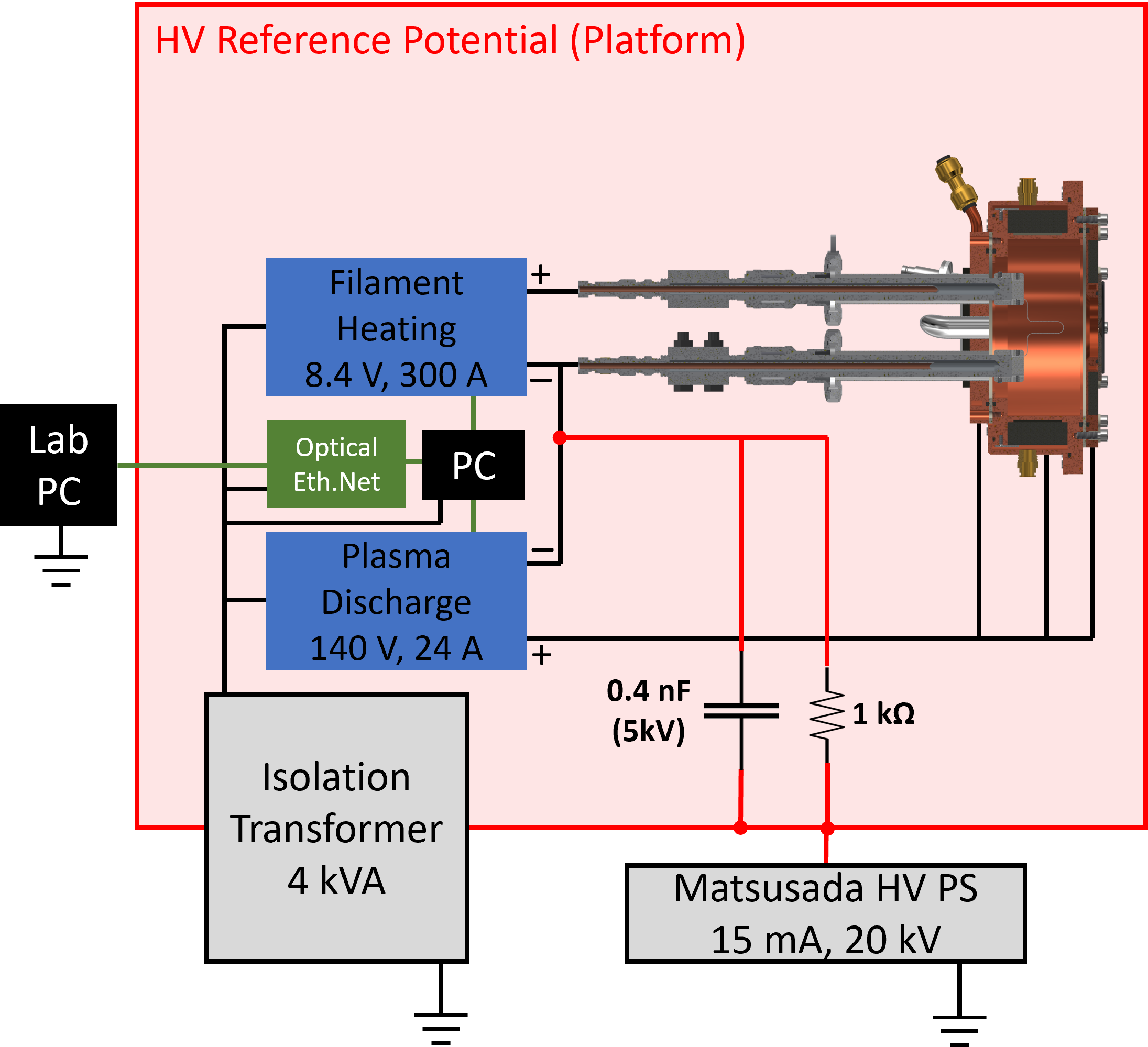}
  \caption{Wiring schematic for the ion source. The red square region corresponds to the HV reference potential. Green lines are data connections. Black lines are power cables. }
  \label{fig:FilamentWiringDiagram}
\end{figure}

The electrical systems are organized into two regions: the high voltage (HV) platform and the analysis region. The source body is located on the HV platform, which can be raised up to 20~kV. Power is transferred to the platform via two 2-kVA transformers. For safety, the entire platform is enclosed in a Faraday cage. Systems directly associated with the source - including the filament power supplies, discharge circuitry, platform computer, and embedded diagnostics - are also located within this Faraday cage in order to shield them from external interference. To further shield these pieces of equipment from interference associated with the HV platform itself, they are enclosed in a smaller Faraday cage within the larger one. A schematic of the platform assembly is shown in FIG.~\ref{fig:FilamentWiringDiagram}.

Outside the Faraday cage, a rack of grounded power supplies provides power to the platform, extraction system, and diagnostics. The main control system, more details in \ref{subsec:ControlSystem}, interfaces with the computer on the HV platform using an optical connection. This computer also controls the hydrogen gas input by communicating with the mass flow controller. This architecture allows for robust, real-time communication while maintaining electrical isolation, enabling safe and stable operation.

Several supplies provide power to different components. Two connect to the filament which is responsible for initiating the \htp plasma. The first raises the temperature of the filament to induce thermionic emission. This is performed by a TDK Lambda (8.4~V / 300~A) power supply. The second maintains a potential difference between the filament and source body to facilitate discharge. This is performed by a Matsusada REK (150~V / 12~A) power supply. These are both located on the platform within the secondary Faraday cage. The platform is raised to high voltage using a Matsusada AU (20 kV / 15 mA) power supply located in the grounded rack. The extraction system, the puller and shaping lenses, are powered by Matsusasda (20~kV / 7.5~mA) power supplies also located in the grounded rack. 

The analysis beamline lies downstream of the extraction system, outside the Faraday cage. The diagnostics are discussed in more detail in \ref{subsec:AnalysisBeamline}. Analysis equipment is powered by additional supplies located in the grounded rack. The Faraday cups which measure beam current are equipped with a suppression electrode which is held at -350 V. The analysis beam also includes dipole and quadrupole magnets which are supplied by independent high-current Lambda and Power Ten supplies respectively. 

The electronics differ modestly between the MIST-1 and MIST-2 systems. The second, smaller Faraday cage within the larger cage enclosing the platform was added to improve robustness and showed a marked improvement in operational stability. We further acquired additional Matsusada power supplies for lenses which previously had been powered by different manufacturers but are now all driven identically. Additionally, to further improve stability, a capacitor was added in parallel to the resistor on the platform, as shown in FIG.~\ref{fig:FilamentWiringDiagram}. 

\subsection{Control System}
\label{subsec:ControlSystem}

For MIST-1, a control system was developed to provide stable semi-automated operation \cite{WEIGEL2023168590}.
It was designed using the open-source React-based framework React Automation Studio (RAS) developed by iThemba Labs \cite{duckitt:cyclotrons2019-tha03, duckitt:icalepcs2023-fr2bco01}.
The backend of the control system is based on EPICS \cite{EPICS}, which enables robust communication between distributed networked systems.
Since many devices in MIST-1 were developed with off-the-shelf microcontrollers such as Arduino and Teensy microcontrollers, a fast and simple communication protocol was developed to integrate EPICS with these devices.

The frontend of the control system is developed using React to provide a modern and easily customizable user interface.
Each instrument in the ion source system is interfaced in a way that allows the read/write values of each device to be displayed and updated frequently.
RAS also provides several features that enable the safe operation of accelerator systems, including alarms.

For devices that read out a substantial amount of data, such as emittance scanners and Faraday cups, we have also implemented new data acquisition methods to efficiently read EPICS process variables and write to disk in highly compressed data formats, like HDF5.

Code for the MIST-1 control system is available freely on our public GitHub repository \cite{mist1-control-system-github}.

\subsection{Analysis beam line}
\label{subsec:AnalysisBeamline}

\begin{figure}[t!]
  \centering
  \includegraphics[width=1.0\linewidth]{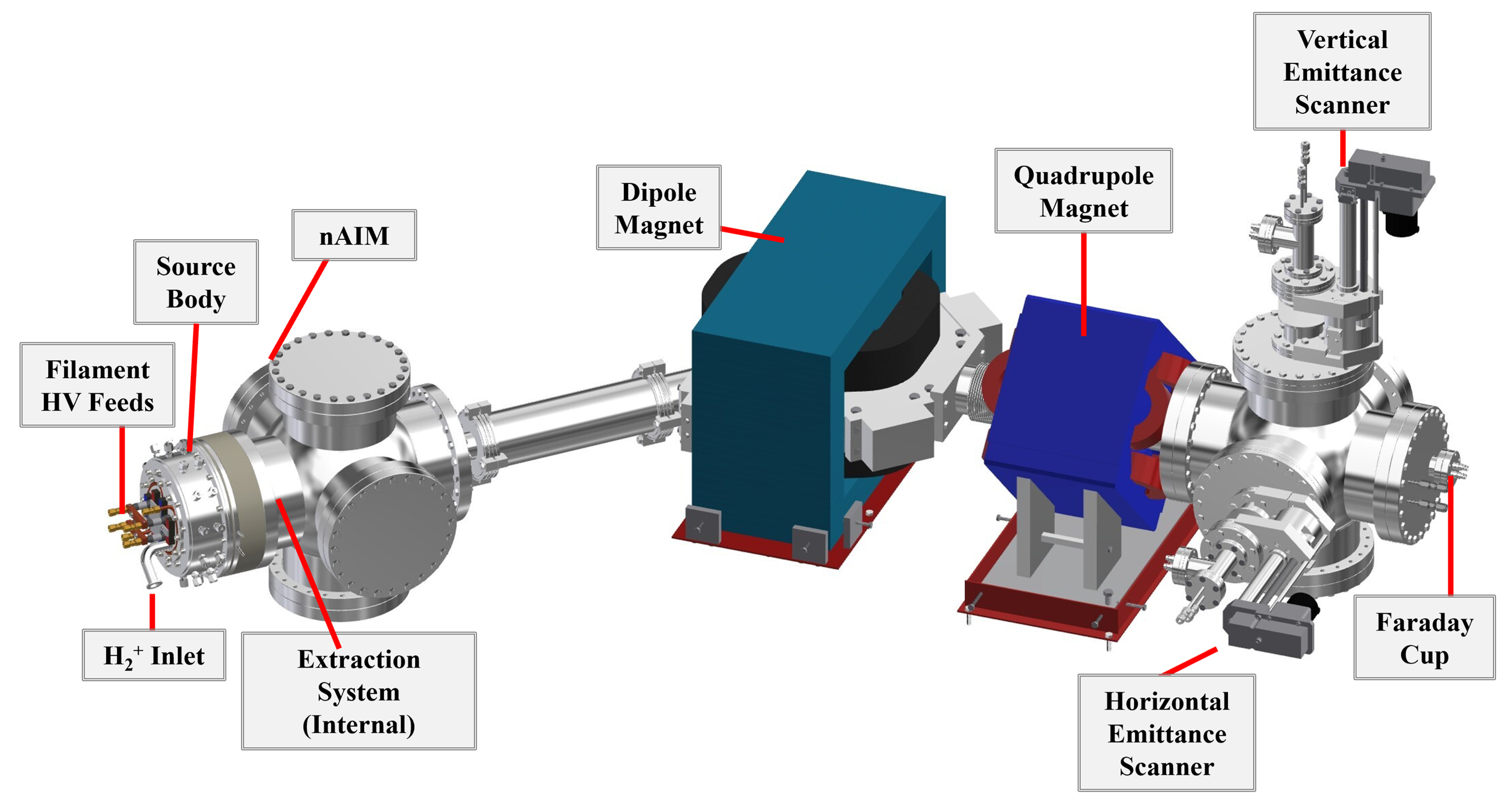}
  \caption{The analysis beam line with diagnostic equipment labeled.}
  \label{fig:AnalysisBeamline}
\end{figure}

After the plasma is extracted from the ion source and focused by the electrostatic lenses, it enters the analysis beamline, a region dedicated to beam characterization. The analysis beamline begins directly following the lenses, inside the first six-way cross. FIGURE~\ref{fig:MIST-2_XS_CAD} illustrates the lens system passing through one of the flanges of this six-way cross. The analysis beamline as a whole is shown in FIG.~\ref{fig:AnalysisBeamline}. 

\begin{table}[b!]
\caption{Equipment use in the analysis beamline.\label{tab:AnalysisBeam} }
\begin{tabular}{|c|c|}
\hline
\textbf{Equipment} & \textbf{Usage} \\
\hline \hline
nAIM & Vacuum monitoring\\
Dipole Magnet & Ion selection by m/q \\
Quadrupole & Vertical or horizontal focusing\\
Allison Scanner (2x) & Vertical \& horizontal emittances \\
Faraday cup & Beam current measurement\\
\hline
\end{tabular}
\end{table}

The analysis section includes a range of instrumentation designed to measure key beam properties such as current, species composition, emittance, and vacuum quality. These diagnostics are essential for evaluating source performance, optimizing extraction and transport parameters, and ensuring compatibility with downstream accelerator components. Virtually all diagnostic components are modular and are either free standing or are built into vacuum flanges which can be attached to any available six-way cross. This provides flexibility for different studies. Table \ref{tab:AnalysisBeam} lists the used diagnostic equipment.

Both the dipole and quadrupole magnets are capable of static focusing, or can have their fields dynamically scanned by the control system in order to take systematic measurements of beam species and focusing, The results of such a study are further discussed in \ref{subsec:MagnetStudy}.

The Faraday cup is read out by a Keithley picoammeter and has a 150~mm x 8~mm slit used to isolate portions of the beam or read out the entire beam current.

\section{Beam simulations\label{sec:simulations}}

\begin{figure}[t!]
  \centering
  \includegraphics[width=1.0\linewidth]{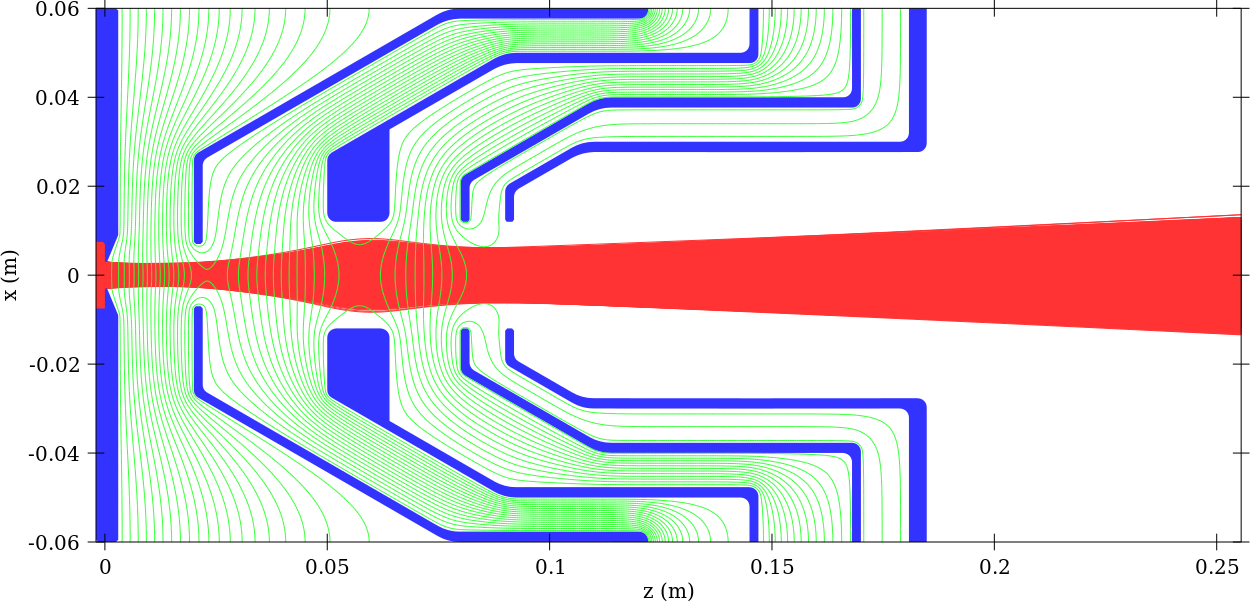}
  \caption{Particle trajectories from an IBSimu simulation of a 10~mA beam going  through the new MIST-2 extraction system, containing
           10\% \hp, 80\% \htp, and 10\% \hthreep.}
  \label{fig:sim_traj}
\end{figure}

\begin{figure}[b!]
  \centering
  \includegraphics[width=1.0\linewidth]{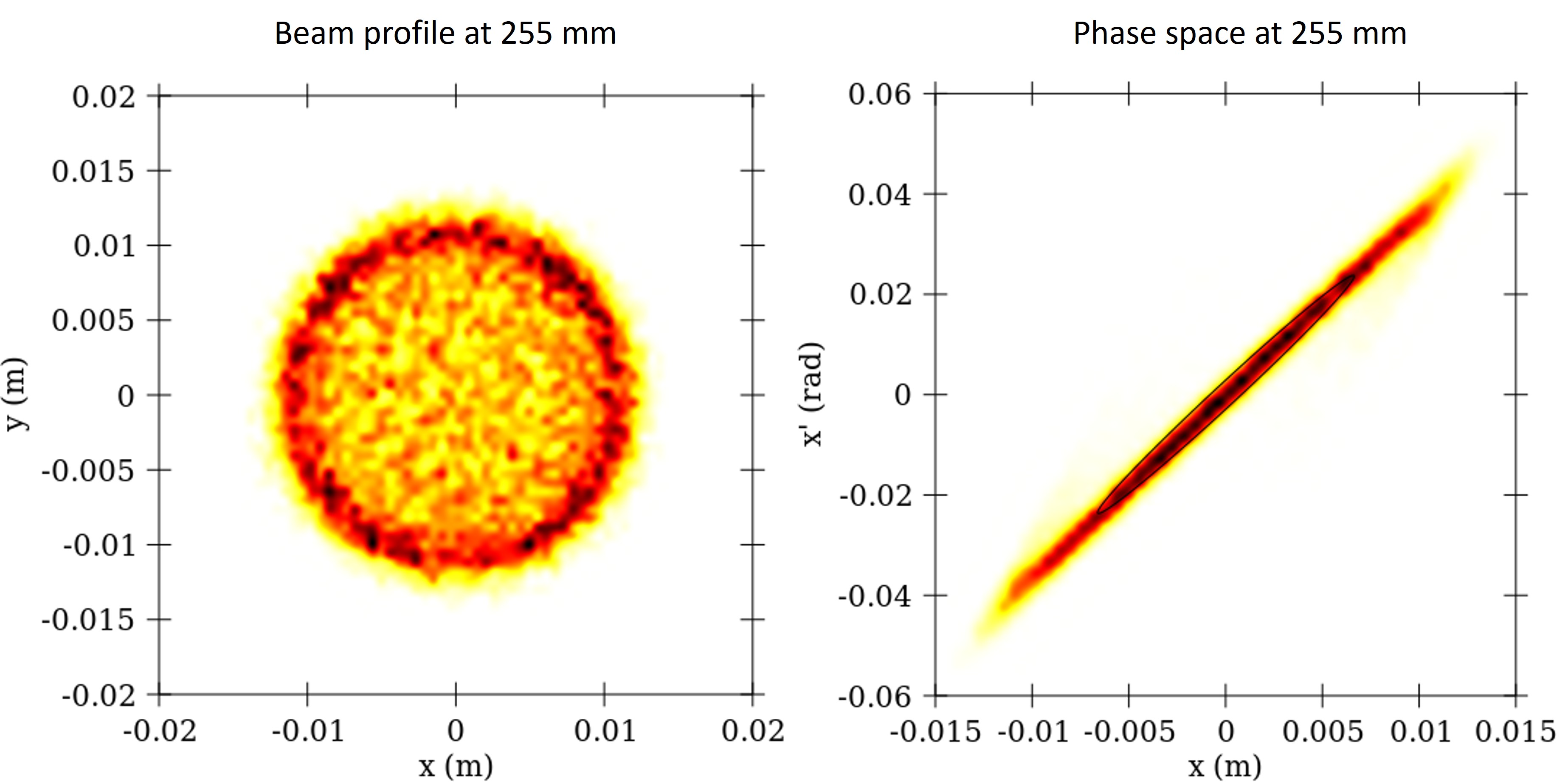}
  \caption{Beam cross-section (left) and xx' phase space (right) 255~mm after the plasma aperture from an IBSimu simulation of a 10~mA beam going through the new MIST-2 extraction system, containing 10\% \hp, 80\% \htp, and 10\% \hthreep.}
  \label{fig:sim_xy_and_xxp}
\end{figure}

For the simulation of beam formation at the ion source plasma aperture,
we used the well-established IBSimu code~\cite{kalvas:ibsimu} which uses a 1D plasma
model to account for the quasi-neutrality of the plasma inside the source and the 
formation of a plasma sheath between plasma and wall. IBSimu uses a 3D iterative stabilized biconjugate gradient method for the calculation of self fields and a Runge-Kutta particle pusher. For the simulations of the transport of the beam through the various iterations of the 
analysis beam line, we used Warp~\cite{friedman:warp}, a particle-in-cell code for high-current beams. Specifically, we used the XY slice solver, which calculates the transverse self-fields at each time step assuming negligible longitudinal self-fields, for efficiency reasons. This is a good approximation, because the beam is a DC beam (a continuous stream of particles) and the changes in transverse size are adiabatic.

In addition to space-charge, we take space charge compensation into account. This is a process by which space charge of the beam is reduced through the accumulation of slow secondary electrons inside the beam envelope. These electrons are produced by ionization of the residual gas molecules by the beam ions. A detailed treatment and semi-empirical model can be found in Ref.~\cite{winklehnerSpaceChargeCompensation2015}. In IBSimu simulations, we assume a conservative compensation factor of 80\% when the beam is well-shielded from external electric fields. In Warp, we use the model proposed in Ref.~\cite{winklehnerSpaceChargeCompensation2015}.

As examples of the IBSimu simulations, we show the beam trajectories in the zx plane and the corresponding cross section and xx' phase space 255~mm after the plasma aperture (location ofthe Faraday cup entrance) for the nominal 10~mA beam operation of the MIST-2 with the new extraction system in FIG.~\ref{fig:sim_traj} and FIG.~\ref{fig:sim_xy_and_xxp}, respectively. The applied voltages are listed in TABLE~\ref{tab:xs_nominal_voltages}. The rms beam diameter is $\sim13.7$~mm, the normalized rms emittance is $\sim0.077$~$\pi$-mm-mrad, and the rms energy spread is $\sim20$~eV in this simulation. We assumed an ion temperature T$_i=2.5$~eV in the plasma.
\begin{figure}[b!]
    \centering
    \includegraphics[width=1.0\columnwidth]{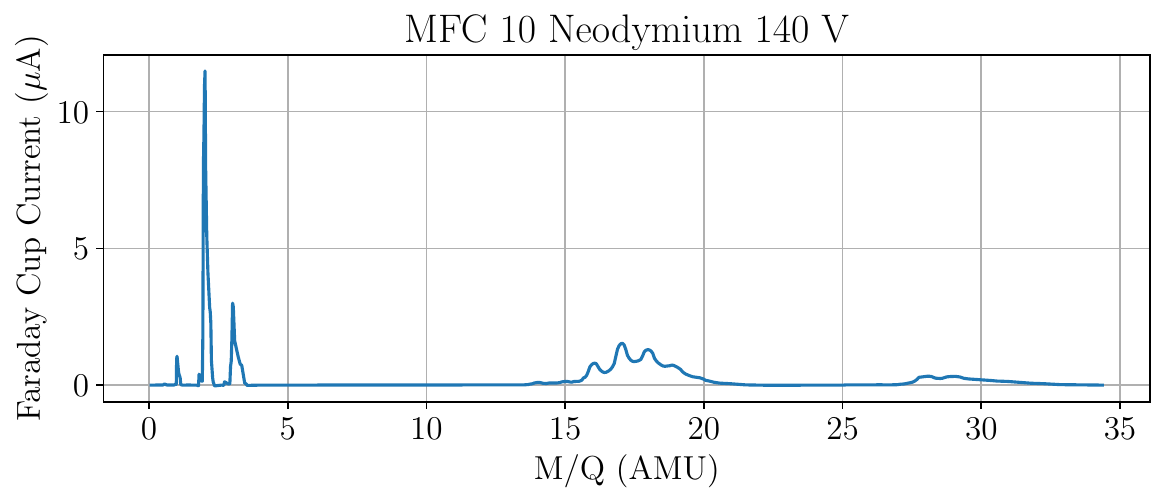}
    \caption{A sample spectrum with Neodymium magnets, the MFC at 10\%, and a discharge voltage of 140~V. The hydrogen peaks are fully resolved, thus we calculate their current by taking their value at the peak. In FIG.~\ref{fig:neodymium-fitted_contams}, we show a zoom of the peaks with M/Q = 14 to 21.}
    \label{fig:neodymium-sample-spectrum}
\end{figure}
For simulations of the analysis beam line, we refer the reader to Refs.~\cite{winklehnerHighcurrentH2Beams2021a, winklehnerNewCommissioningResults2022}. These demonstrated excellent agreement between IBSimu, Warp simulations, and measurements directly after extraction as well as at the end of the analysis beam line.

\section{Measurements}
\label{sec:measurement}

\subsection{Methods}

\begin{figure}
    \centering
    \includegraphics[width=1.0\columnwidth]{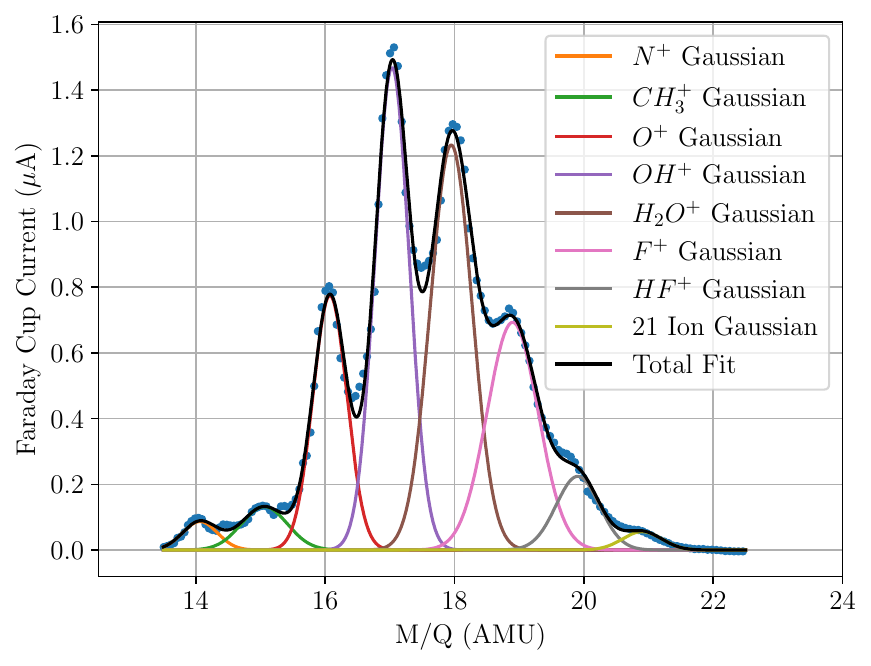}
    \caption{Zoom of FIG.~\ref{fig:neodymium-sample-spectrum} on species with M/Q from 14 to 21 (``first contaminant cluster''). Each peak is fitted with a Gaussian. 
    See TABLE~\ref{tab:fitted-contams} for the fit parameters.}
    \label{fig:neodymium-fitted_contams}
\end{figure}

\begin{table}[b!]
\label{tab:fitted-contams}
\caption{Fit values for the Gaussian curves in FIG.
\ref{fig:neodymium-fitted_contams}}
\begin{tabular}{|c|c|c|c|}
    \hline
    \textbf{Ion Species} & \textbf{Amplitude} & \textbf{Mean} & \textbf{Sigma} \\
    \hline
    \hline
    N$^+$ & 0.087 & 14.06 & 0.269 \\
    CH$_3+$ & 0.132 & 15.06 & 0.354\\
    O$^+$ & 0.775 & 16.07 & 0.264\\
    OH$^+$ & 1.469 & 17.03 & 0.266\\
    H$_2$O$^+$ & 1.233 & 17.95 & 0.323\\
    F$^+$ & 0.693 & 18.90 & 0.395\\
    HF$^+$ & 0.224 & 19.91& 0.336\\
    M/Q$=21$ & 0.056 & 20.90 & 0.301\\
    \hline
    
\end{tabular}
\end{table}

We present two types of measurements here: total currents from the ion source and ion species ratios obtained from mass spectra. For the total currents, we either used a Faraday cup placed in the first diagnostic box or the drain current on the source high voltage power supply, adjusted for dark currents. The latter procedure was verified experimentally in previous experiments~\cite{winklehnerHighcurrentH2Beams2021a}.

All Faraday cups have electron suppression electrodes and we experimentally found the lowest voltage that fully suppresses secondary electrons leaving the cup (-300~V for the first Faraday cup and -500~V for the second, which has a larger aperture).

To analyze the mass spectra, we first converted dipole current to $M/Q$ using a standard quadratic fit. In most spectra, the beam is not fully contained within the 8~mm wide slit of the Faraday cup. However, as we are only reporting ion species fractions and not actual currents and the beam dynamics is identical for each species (all singly-charged) as long as dipole and quadrupole are scaled with a fixed ratio.

For fully resolved peaks, such as the hydrogen species in FIG.~\ref{fig:neodymium-sample-spectrum}, we used the peak height directly. In the case of overlapping peaks, we fitted a sum of Gaussian curves (one for each peak) to the overlapping peaks. We then used the amplitudes of the Gaussians as peak heights. An example is shown in FIG.~\ref{fig:neodymium-fitted_contams}.

Due to the fringe field of the dipole magnet, the beam arriving at the Faraday cup has a triangular cross-section, which causes the asymmetry seen in FIG.~\ref{fig:neodymium-sample-spectrum}. Broadening of higher mass peaks and overlap washes out this effect for the contaminant clusters, making a Gaussian a feasible choice for fit function.


\subsection{Studies with the MIST-2 back plate on the MIST-1 body}
\begin{figure}[!t]
\centering
\includegraphics[width=1.0\linewidth]{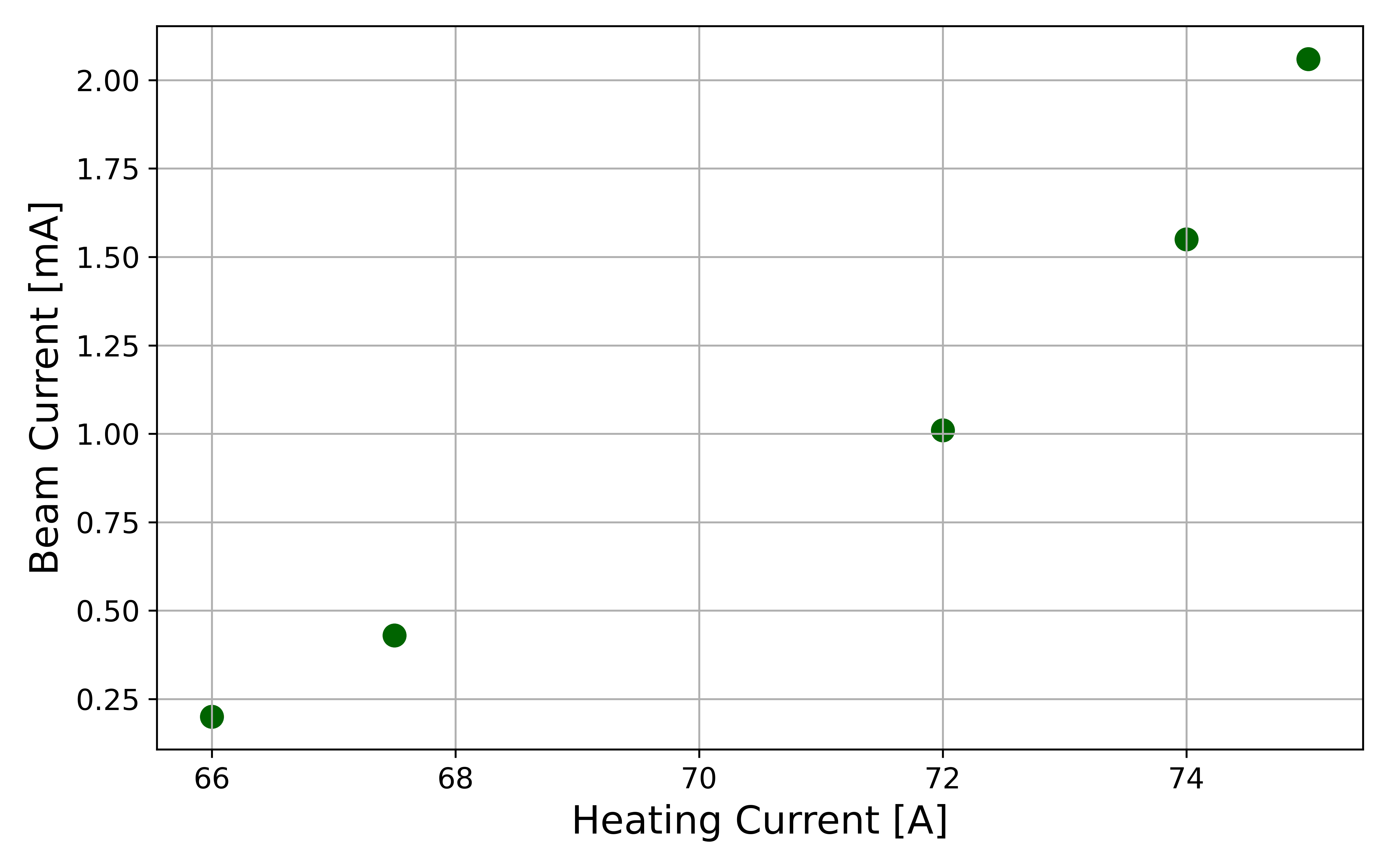}
\caption{Total extracted beam versus filament heating current with the MIST-2 back plate on the MIST-1.}
\label{fig:HeatingStudy}
\end{figure}

\begin{figure}[!b]
\centering
\includegraphics[width=1.0\linewidth]{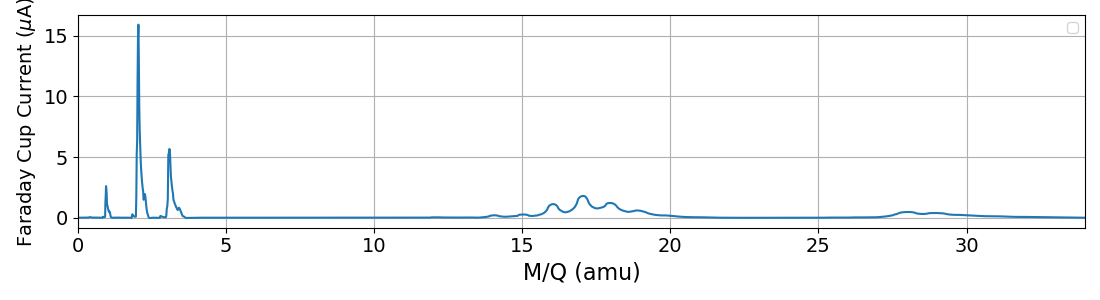}
\caption{Mass spectrum with the MIST-2 backplate. The MFC is set to $10\%$, filament discharge is set to 150V/1.59A, and filament heating is set to 72A. The ion species fractions are \hp = $8.5\%$, \htp = $51.9\%$, \hthreep = $18.5\%$, Other = $21.1\%$.}
\label{fig:Spectrum_Backplate}
\end{figure}

With this intermediate upgrade we were able to test the cooling capacity of the new backplate and the new filament mounting set up. With the source running at 3.4~mA of beam current, the backplate as a whole showed no evidence of heating up. The gas inlet did slightly warm up, but it was not to a concerning extent. Additionally, the tungsten filament was much easier to securely mount with the new back plate and stayed well connected during all of our studies. 

In FIG.~\ref{fig:HeatingStudy} we show total beam current as a function of filament heating current. The rising trend with no sign of flattening out suggests that 
there is room for increased beam current with more heating. 
In FIG.~\ref{fig:Spectrum_Backplate} we show a mass spectrum demonstrating lowered contamination after vacuum baking of the Viton O-rings on the back plate.

\subsection{Permanent magnet study with MIST-1}
\label{subsec:MagnetStudy}
\begin{figure}[t!]
    \centering
    \includegraphics[width=1.0\columnwidth]{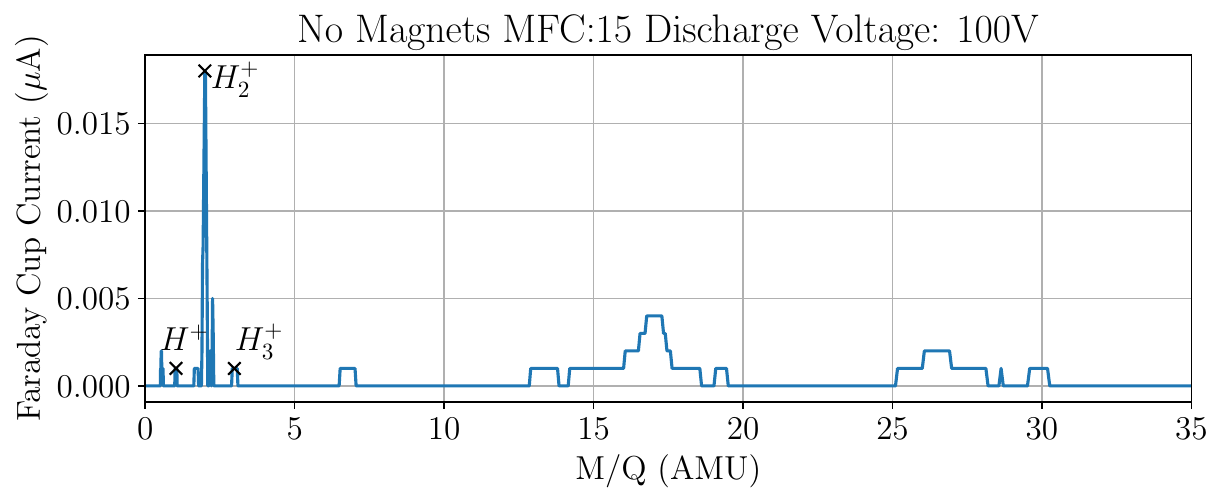}
    \caption{A sample spectrum with no magnets installed in the MIST-1. Here, the \htp fraction was 69.2\% \htp (90\% considering only hydrogen species), while the total extracted current was only 0.026~mA.}
    \label{fig:no-magnet-spectrum}
\end{figure}

\begin{figure*}[tbh]
\centering
\includegraphics[width=1.0\linewidth]{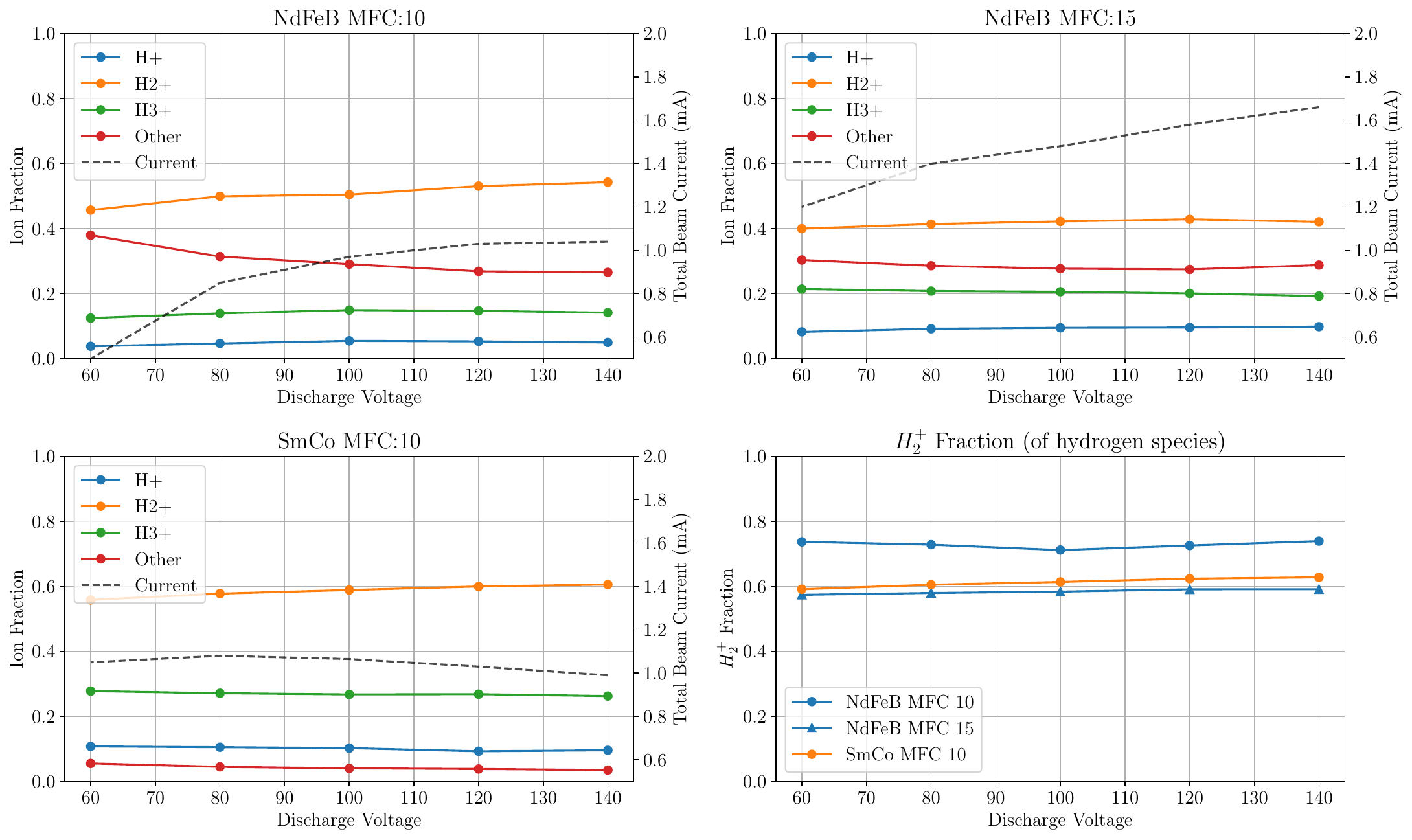}
\caption{Results of the magnet study, showing the ion fractions of each species for each magnet type and MFC setting. The lower right plot compares the \htp fraction for the three different magnet and MFC combinations. Note that it shows the \htp fraction of the hydrogen species, and excludes contaminants, as those are due to non-magnet parameters.}
\label{fig:magnet-study-res}
\end{figure*}

While in the plasma chamber, \htp can dissociate into \hp and recombine to \hthreep. In Ref. \cite{winklehnerHighcurrentH2Beams2021a}, we explored the relationship between ion species fraction and discharge current, discharge voltage, and hydrogen flow. In this study, we sought to understand how the strength of the multicusp magnetic field containing the plasma impacts the ion species fraction and total current, to inform our magnet choice for MIST-2. We mounted two measurement campaigns, one without any magnets and one after replacing the \smco magnets with \ndfe, to compare against our past data (from Ref.~\cite{winklehnerHighcurrentH2Beams2021a}) obtained with \smco magnets.

With no magnets, and thus no containment, we saw very little total beam current. With the MFC at 40\% and a discharge voltage of 140~V, we observed a maximum total current without magnets of 0.11~mA. In addition, at lower discharge voltages, we often struggled to ignite a plasma at all. Only with higher settings on the MFC and high discharge voltages could we extract beam. With too few data points in the same MFC and discharge regime, we are not plotting the ``no magnet'' results together with the two permanent magnet types. However, we did observe the highest \htp fraction (see FIG. \ref{fig:no-magnet-spectrum}) to date at 90\% \htp (considering only hydrogen species--69.2\% with contaminants), albeit at a low total extracted current of 0.026~mA.

We performed scans of the discharge voltage with the neodymium magnets with the MFC at 10\% and at 15\% to compare with the set of measurements we had taken in Ref.~\cite{winklehnerHighcurrentH2Beams2021a} with samarium-cobalt magnets.
In FIG. \ref{fig:magnet-study-res}, we show the ion fractions of each magnet type and MFC setting combination versus discharge voltages, as well as a comparison of the \htp fraction across the three settings. Because the contaminants varied significantly, the \htp fraction comparison considers only the hydrogen peaks. 
The neodymium measurements have higher contaminant peaks. We believe this is due to vacuum conditions and outgassing of O-rings and have identified a path forward to reduce the contaminants further.
In addition to displaying the fractional results, FIG.~\ref{fig:magnet-study-res} corroborates the findings in Ref.~\cite{winklehnerHighcurrentH2Beams2021a} that increased gas flow leads to a higher \hthreep fraction and that discharge voltage seems to have a minimal effect on ion species fractions.

\begin{figure}[tbh]
\centering
\includegraphics[width=1.0\linewidth]{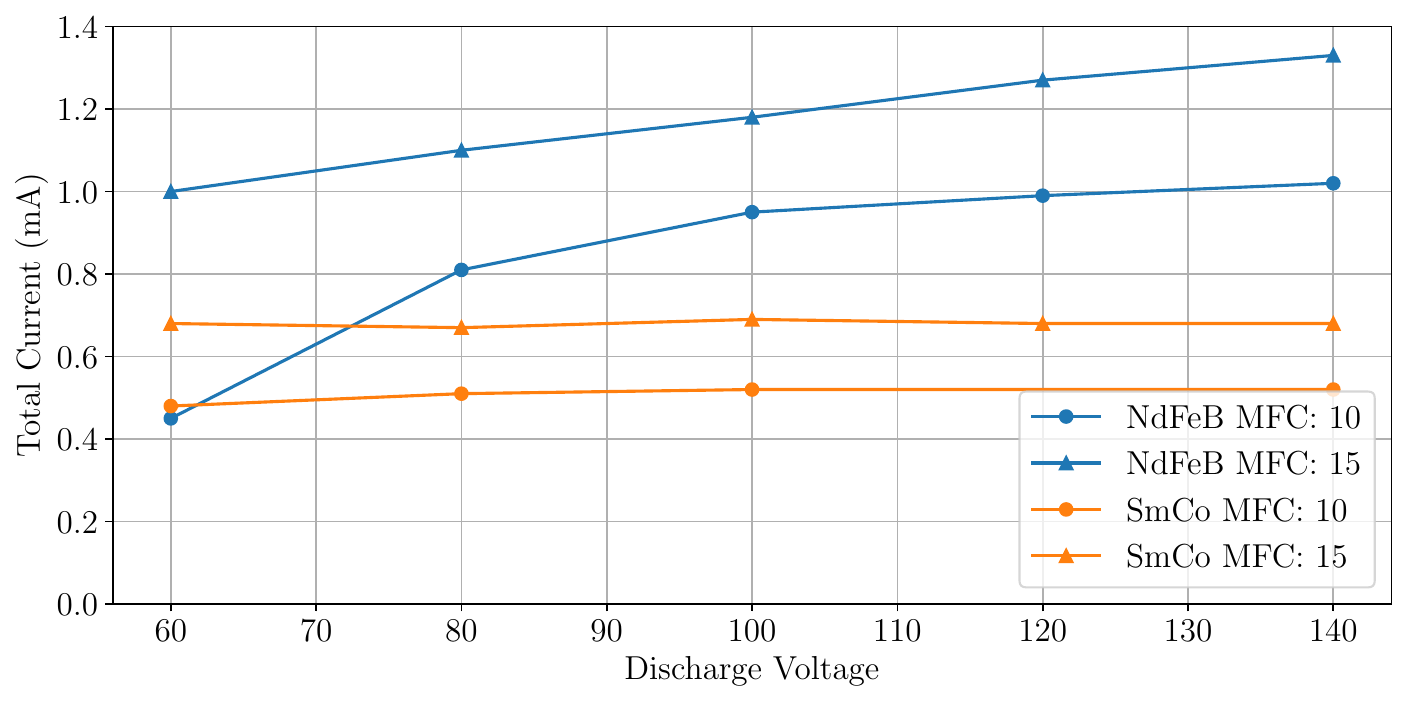}
\caption{Comparison of total beam current versus discharge voltage for \ndfe and \smco magnets at MFC settings 10\% and 15\%. This set of measurements was done independently of the ones reported in FIG.~\ref{fig:magnet-study-res}}. 
\label{fig:current-investigation}
\end{figure}

We also investigated how the total current changes with discharge voltage under identical source parameters. The results can be seen in FIG. \ref{fig:current-investigation}. The neodymium magnets give higher total current than the \smco magnets.

We have elected to use neodymium magnets for \mbox{MIST-2} due to their higher \htp fraction and higher total current.

\subsection{First commissioning results with MIST-2}
\begin{figure}[!t]
\centering
\includegraphics[width=1.0\linewidth]{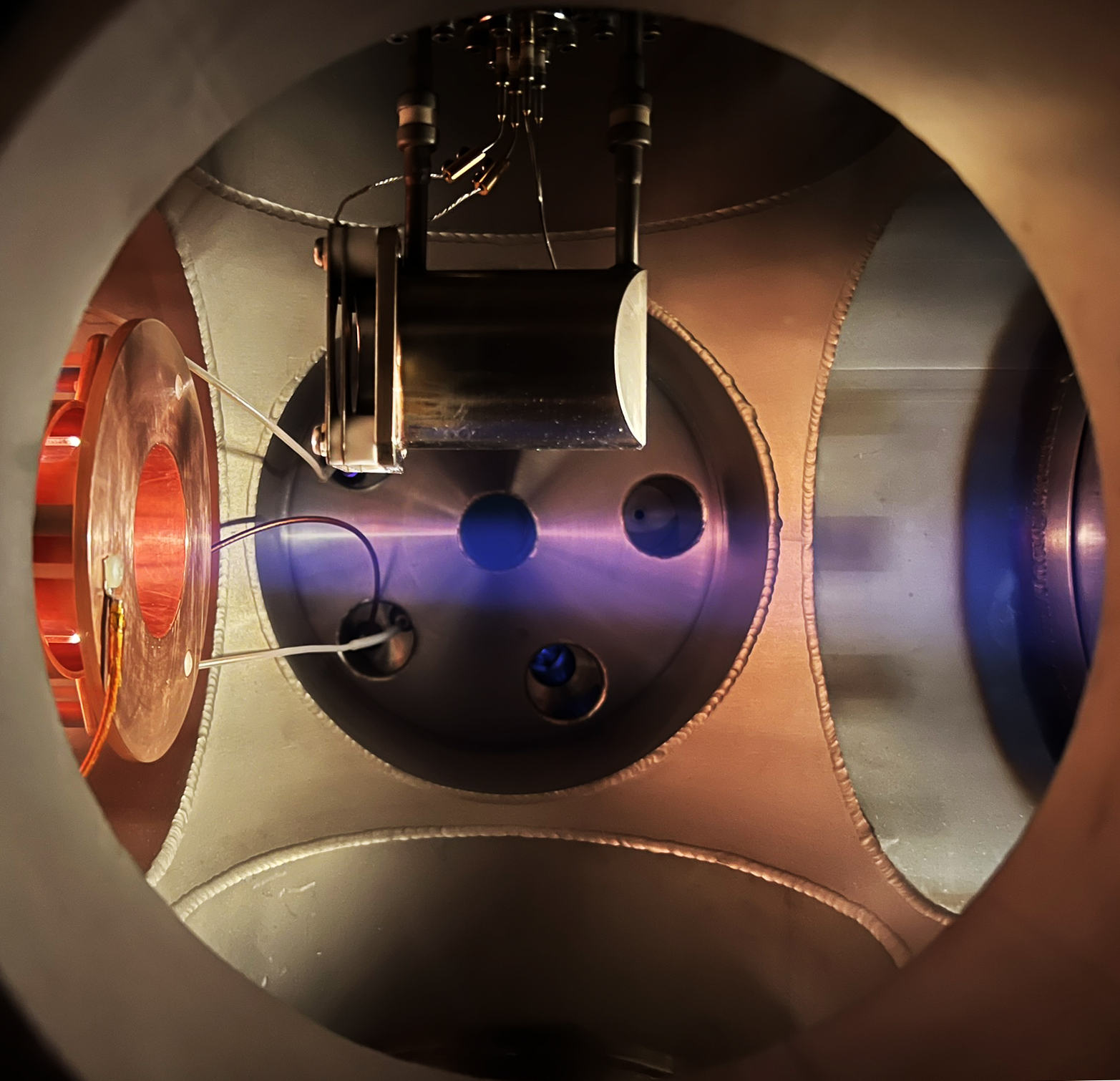}
\caption{Photograph of a high-current beam exiting the extraction system and passing underneath
         the retracted Faraday cup. The atomic interactions of the beam ions with the 
         residual gas lead to the characteristic purple glow.}
\label{fig:beam_photo}
\end{figure}
After dismantling the MIST-1 ion source and rebuilding the high voltage platform and
first diagnostic box, we assembled the MIST-2 ion source and performed the first 
commissioning tests. With a base pressure of $2.2\cdot10^{-7}$~Torr, we performed high voltage conditioning up to the nominal values listed in TABLE~\ref{tab:xs_nominal_voltages} within an hour, and then repeated the process at $4.5\cdot10^{-6}$~Torr.\\

We then proceeded to increase the filament heating until we reached a plasma discharge current of 20~A (at 140~V) and a total current
of 6~mA in the Faraday cup installed right after the extraction system. We show a photograph of the 6~mA beam passing underneath the retracted Faraday cup
in FIG.~\ref{fig:beam_photo}. The purple glow stems from interactions of the beam ions with
the residual gas, predominantly H$_2$. A slight offset in the vertical direction is caused by
a very small misalignment of the extraction system with the plasma aperture. Image analysis of
the photograph shows that the beam size is in excellent agreement with the IBSimu simulations
reported in Section~\ref{sec:simulations}. The simulated normalized rms emittance at the entrance of the 
Faraday cup (255~mm after the plasma aperture) is 0.079~$\pi$-mm-mrad for a 6~mA beam. Based on previous comparisons of IBSimu and measured emittances for the MIST-1~\cite{winklehnerHighcurrentH2Beams2021a}, we trust this number.

The commissioning goal of this first run was to demonstrate stable operation at high current, which we reached by running for one hour at 6~mA. We increased the heating current more and 
were able to get to 6.8~mA of beam in the Faraday cup, however, at that point, we were at
the current limit of 22~A of the discharge power supply and roughly every 5~min, we would 
experience a over-current protection trip from the power supply.

During the entire operation, the ion source remained close to room temperature, with the exception of the gas feed tube, which is made from stainless steel to facilitate the vacuum brazing into the back plate. Due to its location right behind the hot filament, its temperature reached about 45\degree~C from radiative heating. This is within acceptable parameters.

\section{Outlook}

\subsection{Perspectives in Machine Learning}
The development of next-generation accelerator systems like the MIST ion sources require extreme care in both their design and their deployment. Experimental idiosyncrasies only complicate the issue further; for example, IsoDAR's planned installation at the Yemilab Center for Underground Physics enforces strict constrains on the size and shape of its larger components. The collaboration's interest in Machine Learning (ML) has resulted in successful development of particle-in-cell surrogate models which can quickly simulate beam throughput \cite{koserInputBeamMatching2022},  significantly speed up multi-objective design optimization \cite{villarrealNeuralNetworksEffective2024}, and validate IsoDAR's cyclotron design robustness \cite{winklehnerOrdermagnitudeBeamCurrent2022} through uncertainty quantification \cite{doi:10.1137/16M1061928}.

While ML has found substantive use in early phases of accelerator design, real-time commissioning and maintenance are typically left to human drivers. As accelerator stacks become more complicated, however, seamless operation becomes more difficult. Reinforcement learning strategies are beginning to gain traction in autonomous real-time fine-tuning of accelerator systems~\cite{Kaiser2024-xw}, though ever-evolving laboratory conditions may necessitate the inclusion of more modern continual learning techniques~\cite{rajput2025outlookdeployablecontinuallearning}. 

MIST-2 presents a fertile test-bed for the development of autonomous operation agents and ML-powered co-pilots. Its eventual linkage to the RFQ, spiral inflector, and HCHC-XX cyclotron require precise maintenance of the quantity and quality of its extracted \htp beam and of course the operation of the system as a whole. Further, the EPICS control system provides a natural framework for real-world training data generation and easily integrable automation software, while its hysteretic dependence on its design settings present a host of opportune challenges for the development of deployable ML suites which can find their use in assisting technicians at accelerators of all kinds.

In preparation for the development of an ML co-pilot operator, we have developed software which automatically runs the ion source with user-specified varying design parameters subject to appropriate laboratory conditions and physically realizable settings. The control system is extended to be able to read and execute a list source settings contained in a generated configuration file. A combination of real-world data gathered by the sampler and training data generated by simulators provide an important first step for a human-AI driving cooperative. 

\subsection{Further Outlook}

Currently, the MIST-2 ion source is in storage awaiting the installation of water and
power in a refurbished laboratory space. When we reassemble the source, we plan the following upgrades.

\begin{itemize}
\setlength\itemsep{0.1em}
    \item Redesign of the analysis beam line. In past measurements, we have observed low transmission efficiency through the analysis beam line. While this does not change the results we have shown here, higher transmission means higher statistics for mass spectra and a better test bed for high current beam handling.
    \item Procurement of a higher power discharge power supply. The commissioning of MIST-2 has shown that we operate at both voltage and current limits of our power supply while other components (e.g., filament heating) could go much higher still. 
    \item Addition of a Langmuir probe. We made provisions for this on the back plate and have a mechanical design and a circuit design ready.
    \item Significant quality-of-life improvements by using AI/ML, as outlined in the previous section, for stabilizing and auto-tuning.
    \item We are investigating building a retarding field analyzer to measure the 
    energy distribution of the beam.
\end{itemize}

We expect a significant performance boost from these improvements, as well as improved analysis capabilities.

\section{Conclusion}
\label{sec:conclusion}
In this paper, we presented the current status of the multicusp ion source development
at MIT. While we primarily develop these sources to fulfill the stringent requirements
of the IsoDAR project and the HCHC-XX cyclotron design, we believe that high-current,
high-purity \htp sources that produce low-emittance beams might be of interest to other projects as well.
We described in detail the mechanical and electrical design of our latest ion source, the MIST-2, which improves over the MIST-1 in several aspects and has already demonstrated a factor of 2 increase in total current at almost 7~mA of total extracted beam current. From previous experience, we estimate an \htp fraction of about 60\% for this beam. We also showed a test of different permanent magnet materials (including no magnets at all). The result of this study was that no magnets, while giving a very high \htp fraction, do not provide adequate beam current. Between \smco and \ndfe magnets, the \ndfe magnets seem to be performing better in total beam (due to the higher plasma density) and also in \htp fraction. Hence we decided to continue MIST-2 operation with \ndfe magnets.
With a stronger discharge power supply and increasing the filament heating, we expect a further increase in performance by 80\% to reach the nominal beam current requirements of the HCHC-XX.


%
%

%
\newpage
\begin{acknowledgments}
The authors would like to thank Olli Tarvainen for the suggestion to measure 
the ion species distribution without any magnets and the MIT Plasma Science and Fusion
Center for their generous facility support.
This work was supported by the U.S. Department of Energy under 
awards DE-SC0024914 and DE-SC0024138 and the 
U.S. National Science Foundation under award PHY-2411745.
DW was also supported by the Heising-Simons Foundation. 
EW thanks the MIT UROP Office.
JV is supported by the National
Science Foundation Graduate Research Fellowship under
Grant No. 2141064.
\end{acknowledgments}

\section*{Data Availability Statement}
Data is available from the authors upon reasonable request.

\bibliography{mist2}

\begin{thebibliography}{43}%
\makeatletter
\providecommand \@ifxundefined [1]{%
 \@ifx{#1\undefined}
}%
\providecommand \@ifnum [1]{%
 \ifnum #1\expandafter \@firstoftwo
 \else \expandafter \@secondoftwo
 \fi
}%
\providecommand \@ifx [1]{%
 \ifx #1\expandafter \@firstoftwo
 \else \expandafter \@secondoftwo
 \fi
}%
\providecommand \natexlab [1]{#1}%
\providecommand \enquote  [1]{``#1''}%
\providecommand \bibnamefont  [1]{#1}%
\providecommand \bibfnamefont [1]{#1}%
\providecommand \citenamefont [1]{#1}%
\providecommand \href@noop [0]{\@secondoftwo}%
\providecommand \href [0]{\begingroup \@sanitize@url \@href}%
\providecommand \@href[1]{\@@startlink{#1}\@@href}%
\providecommand \@@href[1]{\endgroup#1\@@endlink}%
\providecommand \@sanitize@url [0]{\catcode `\\12\catcode `\$12\catcode `\&12\catcode `\#12\catcode `\^12\catcode `\_12\catcode `\%12\relax}%
\providecommand \@@startlink[1]{}%
\providecommand \@@endlink[0]{}%
\providecommand \url  [0]{\begingroup\@sanitize@url \@url }%
\providecommand \@url [1]{\endgroup\@href {#1}{\urlprefix }}%
\providecommand \urlprefix  [0]{URL }%
\providecommand \Eprint [0]{\href }%
\providecommand \doibase [0]{http://dx.doi.org/}%
\providecommand \selectlanguage [0]{\@gobble}%
\providecommand \bibinfo  [0]{\@secondoftwo}%
\providecommand \bibfield  [0]{\@secondoftwo}%
\providecommand \translation [1]{[#1]}%
\providecommand \BibitemOpen [0]{}%
\providecommand \bibitemStop [0]{}%
\providecommand \bibitemNoStop [0]{.\EOS\space}%
\providecommand \EOS [0]{\spacefactor3000\relax}%
\providecommand \BibitemShut  [1]{\csname bibitem#1\endcsname}%
\let\auto@bib@innerbib\@empty
\bibitem [{\citenamefont {Conrad}\ \emph {et~al.}(2014)\citenamefont {Conrad}, \citenamefont {Shaevitz}, \citenamefont {Shimizu}, \citenamefont {Spitz}, \citenamefont {Toups},\ and\ \citenamefont {Winslow}}]{conradPrecisionAntineutrinoelectronScattering2014}%
  \BibitemOpen
  \bibfield  {author} {\bibinfo {author} {\bibfnamefont {J.~M.}\ \bibnamefont {Conrad}}, \bibinfo {author} {\bibfnamefont {M.~H.}\ \bibnamefont {Shaevitz}}, \bibinfo {author} {\bibfnamefont {I.}~\bibnamefont {Shimizu}}, \bibinfo {author} {\bibfnamefont {J.}~\bibnamefont {Spitz}}, \bibinfo {author} {\bibfnamefont {M.}~\bibnamefont {Toups}}, \ and\ \bibinfo {author} {\bibfnamefont {L.}~\bibnamefont {Winslow}},\ }\href {\doibase 10.1103/PhysRevD.89.072010} {\bibfield  {journal} {\bibinfo  {journal} {Physical Review D}\ }\textbf {\bibinfo {volume} {89}},\ \bibinfo {pages} {072010} (\bibinfo {year} {2014})}\BibitemShut {NoStop}%
\bibitem [{\citenamefont {Alonso}\ \emph {et~al.}(2022)\citenamefont {Alonso}, \citenamefont {Arg{\"u}elles}, \citenamefont {Bungau}, \citenamefont {Conrad}, \citenamefont {Dutta}, \citenamefont {Kim}, \citenamefont {Marzec}, \citenamefont {Mishins}, \citenamefont {Seo}, \citenamefont {Shaevitz}, \citenamefont {Spitz}, \citenamefont {Thompson}, \citenamefont {Waites},\ and\ \citenamefont {Winklehner}}]{alonsoNeutrinoPhysicsOpportunities2022}%
  \BibitemOpen
  \bibfield  {author} {\bibinfo {author} {\bibfnamefont {J.}~\bibnamefont {Alonso}}, \bibinfo {author} {\bibfnamefont {C.~A.}\ \bibnamefont {Arg{\"u}elles}}, \bibinfo {author} {\bibfnamefont {A.}~\bibnamefont {Bungau}}, \bibinfo {author} {\bibfnamefont {J.~M.}\ \bibnamefont {Conrad}}, \bibinfo {author} {\bibfnamefont {B.}~\bibnamefont {Dutta}}, \bibinfo {author} {\bibfnamefont {Y.~D.}\ \bibnamefont {Kim}}, \bibinfo {author} {\bibfnamefont {E.}~\bibnamefont {Marzec}}, \bibinfo {author} {\bibfnamefont {D.}~\bibnamefont {Mishins}}, \bibinfo {author} {\bibfnamefont {S.~H.}\ \bibnamefont {Seo}}, \bibinfo {author} {\bibfnamefont {M.}~\bibnamefont {Shaevitz}}, \bibinfo {author} {\bibfnamefont {J.}~\bibnamefont {Spitz}}, \bibinfo {author} {\bibfnamefont {A.}~\bibnamefont {Thompson}}, \bibinfo {author} {\bibfnamefont {L.}~\bibnamefont {Waites}}, \ and\ \bibinfo {author} {\bibfnamefont {D.}~\bibnamefont {Winklehner}},\ }\href {\doibase 10.1103/PhysRevD.105.052009} {\bibfield  {journal} {\bibinfo  {journal} {Physical
  Review D}\ }\textbf {\bibinfo {volume} {105}},\ \bibinfo {pages} {052009} (\bibinfo {year} {2022})}\BibitemShut {NoStop}%
\bibitem [{\citenamefont {Alonso}\ \emph {et~al.}(2021)\citenamefont {Alonso}, \citenamefont {Bang}, \citenamefont {Barlow}, \citenamefont {Bartoszek}, \citenamefont {Bungau}, \citenamefont {Calabretta}, \citenamefont {Conrad}, \citenamefont {Kayser}, \citenamefont {Kim}, \citenamefont {Park}, \citenamefont {Seo}, \citenamefont {Shaevitz}, \citenamefont {Spitz}, \citenamefont {Waites},\ and\ \citenamefont {Winklehner}}]{alonsoIsoDARYemilabConceptualDesign2021}%
  \BibitemOpen
  \bibfield  {author} {\bibinfo {author} {\bibfnamefont {J.~R.}\ \bibnamefont {Alonso}}, \bibinfo {author} {\bibfnamefont {K.~M.}\ \bibnamefont {Bang}}, \bibinfo {author} {\bibfnamefont {R.}~\bibnamefont {Barlow}}, \bibinfo {author} {\bibfnamefont {L.}~\bibnamefont {Bartoszek}}, \bibinfo {author} {\bibfnamefont {A.}~\bibnamefont {Bungau}}, \bibinfo {author} {\bibfnamefont {L.}~\bibnamefont {Calabretta}}, \bibinfo {author} {\bibfnamefont {J.~M.}\ \bibnamefont {Conrad}}, \bibinfo {author} {\bibfnamefont {S.}~\bibnamefont {Kayser}}, \bibinfo {author} {\bibfnamefont {Y.~D.}\ \bibnamefont {Kim}}, \bibinfo {author} {\bibfnamefont {K.~S.}\ \bibnamefont {Park}}, \bibinfo {author} {\bibfnamefont {S.~H.}\ \bibnamefont {Seo}}, \bibinfo {author} {\bibfnamefont {M.~H.}\ \bibnamefont {Shaevitz}}, \bibinfo {author} {\bibfnamefont {J.}~\bibnamefont {Spitz}}, \bibinfo {author} {\bibfnamefont {L.~H.}\ \bibnamefont {Waites}}, \ and\ \bibinfo {author} {\bibfnamefont {D.}~\bibnamefont {Winklehner}},\ }\href {\doibase
  10.48550/arXiv.2110.10635} {\enquote {\bibinfo {title} {{{IsoDAR}}@{{Yemilab}}: {{A Conceptual Design Report}} for the {{Deployment}} of the {{Isotope Decay-At-Rest Experiment}} in {{Korea}}'s {{New Underground Laboratory}}, {{Yemilab}}},}\ } (\bibinfo {year} {2021}),\ \Eprint {http://arxiv.org/abs/2110.10635} {arXiv:2110.10635 [hep-ex, physics:physics]} \BibitemShut {NoStop}%
\bibitem [{\citenamefont {Winklehner}(2023)}]{winklehnerIsoDARYemilabADefinitiveSearch2023}%
  \BibitemOpen
  \bibfield  {author} {\bibinfo {author} {\bibfnamefont {D.}~\bibnamefont {Winklehner}},\ }\href {\doibase 10.3390/psf2023008021} {\bibfield  {journal} {\bibinfo  {journal} {Physical Sciences Forum}\ }\textbf {\bibinfo {volume} {8}},\ \bibinfo {pages} {21} (\bibinfo {year} {2023})}\BibitemShut {NoStop}%
\bibitem [{\citenamefont {Conrad}\ \emph {et~al.}(2013)\citenamefont {Conrad}, \citenamefont {Louis},\ and\ \citenamefont {Shaevitz}}]{conradSterileNeutrinos}%
  \BibitemOpen
  \bibfield  {author} {\bibinfo {author} {\bibfnamefont {J.~M.}\ \bibnamefont {Conrad}}, \bibinfo {author} {\bibfnamefont {W.~C.}\ \bibnamefont {Louis}}, \ and\ \bibinfo {author} {\bibfnamefont {M.~H.}\ \bibnamefont {Shaevitz}},\ }\href {\doibase 10.1146/annurev-nucl-102711-094957} {\bibfield  {journal} {\bibinfo  {journal} {Annual Reviews}\ }\textbf {\bibinfo {volume} {63:}},\ \bibinfo {pages} {45} (\bibinfo {year} {2013})}\BibitemShut {NoStop}%
\bibitem [{\citenamefont {Bungau}\ \emph {et~al.}(2024)\citenamefont {Bungau}, \citenamefont {Alonso}, \citenamefont {Barlow}, \citenamefont {Bartozsek}, \citenamefont {Conrad}, \citenamefont {Shaevitz}, \citenamefont {Spitz},\ and\ \citenamefont {Winklehner}}]{bungauNeutrinoYieldNeutron2024}%
  \BibitemOpen
  \bibfield  {author} {\bibinfo {author} {\bibfnamefont {A.}~\bibnamefont {Bungau}}, \bibinfo {author} {\bibfnamefont {J.}~\bibnamefont {Alonso}}, \bibinfo {author} {\bibfnamefont {R.}~\bibnamefont {Barlow}}, \bibinfo {author} {\bibfnamefont {L.}~\bibnamefont {Bartozsek}}, \bibinfo {author} {\bibfnamefont {J.}~\bibnamefont {Conrad}}, \bibinfo {author} {\bibfnamefont {M.}~\bibnamefont {Shaevitz}}, \bibinfo {author} {\bibfnamefont {J.}~\bibnamefont {Spitz}}, \ and\ \bibinfo {author} {\bibfnamefont {D.}~\bibnamefont {Winklehner}},\ }\href {\doibase 10.48550/arXiv.2409.10211} {\enquote {\bibinfo {title} {Neutrino yield and neutron shielding calculations for a high-power target installed in an underground setting},}\ } (\bibinfo {year} {2024}),\ \Eprint {http://arxiv.org/abs/2409.10211} {arXiv:2409.10211 [hep-ex]} \BibitemShut {NoStop}%
\bibitem [{\citenamefont {Winklehner}\ \emph {et~al.}(2024{\natexlab{a}})\citenamefont {Winklehner}, \citenamefont {Spitz}, \citenamefont {Alonso}, \citenamefont {Conrad}, \citenamefont {Moon}, \citenamefont {Herrod}, \citenamefont {De~Neuter}, \citenamefont {Forton}, \citenamefont {Joassin}, \citenamefont {{Van der Kraaij}},\ and\ \citenamefont {W{\'e}ry}}]{winklehnerIsoDARYemilabPreliminaryDesign2024}%
  \BibitemOpen
  \bibfield  {author} {\bibinfo {author} {\bibfnamefont {D.}~\bibnamefont {Winklehner}}, \bibinfo {author} {\bibfnamefont {J.}~\bibnamefont {Spitz}}, \bibinfo {author} {\bibfnamefont {J.~R.}\ \bibnamefont {Alonso}}, \bibinfo {author} {\bibfnamefont {J.~M.}\ \bibnamefont {Conrad}}, \bibinfo {author} {\bibfnamefont {J.}~\bibnamefont {Moon}}, \bibinfo {author} {\bibfnamefont {A.}~\bibnamefont {Herrod}}, \bibinfo {author} {\bibfnamefont {S.}~\bibnamefont {De~Neuter}}, \bibinfo {author} {\bibfnamefont {E.}~\bibnamefont {Forton}}, \bibinfo {author} {\bibfnamefont {D.}~\bibnamefont {Joassin}}, \bibinfo {author} {\bibfnamefont {E.}~\bibnamefont {{Van der Kraaij}}}, \ and\ \bibinfo {author} {\bibfnamefont {G.}~\bibnamefont {W{\'e}ry}},\ }\href@noop {} {\enquote {\bibinfo {title} {{{IsoDAR}}@{{Yemilab}}: {{Preliminary Design Report}} -- {{Volume I}}: {{Cyclotron Driver}}},}\ } (\bibinfo {year} {2024}{\natexlab{a}}),\ \Eprint {http://arxiv.org/abs/2404.06281} {arXiv:2404.06281 [hep-ex, physics:physics]} \BibitemShut
  {NoStop}%
\bibitem [{\citenamefont {Winklehner}\ \emph {et~al.}(2021{\natexlab{a}})\citenamefont {Winklehner}, \citenamefont {Conrad}, \citenamefont {Koser}, \citenamefont {Smolsky},\ and\ \citenamefont {Waites}}]{winklehnerHighCurrentH2Beams2021}%
  \BibitemOpen
  \bibfield  {author} {\bibinfo {author} {\bibfnamefont {D.}~\bibnamefont {Winklehner}}, \bibinfo {author} {\bibfnamefont {J.}~\bibnamefont {Conrad}}, \bibinfo {author} {\bibfnamefont {D.}~\bibnamefont {Koser}}, \bibinfo {author} {\bibfnamefont {J.}~\bibnamefont {Smolsky}}, \ and\ \bibinfo {author} {\bibfnamefont {L.}~\bibnamefont {Waites}},\ }in\ \href {\doibase 10.18429/JACOW-IPAC2021-TUXB07} {\emph {\bibinfo {booktitle} {Proceedings of the 12th {{International Particle Accelerator Conference}}}}},\ Vol.\ \bibinfo {volume} {IPAC2021}\ (\bibinfo  {publisher} {JACoW Publishing, Geneva, Switzerland},\ \bibinfo {year} {2021})\ pp.\ \bibinfo {pages} {4 pages, 0.582 MB}\BibitemShut {NoStop}%
\bibitem [{\citenamefont {H{\"o}ltermann}\ \emph {et~al.}(2021)\citenamefont {H{\"o}ltermann}, \citenamefont {Conrad}, \citenamefont {Koser}, \citenamefont {Koubek}, \citenamefont {Podlech}, \citenamefont {Ratzinger}, \citenamefont {Schuett}, \citenamefont {Smolsky}, \citenamefont {Syha}, \citenamefont {Waites},\ and\ \citenamefont {Winklehner}}]{holtermannTechnicalDesignRFQ2021}%
  \BibitemOpen
  \bibfield  {author} {\bibinfo {author} {\bibfnamefont {H.}~\bibnamefont {H{\"o}ltermann}}, \bibinfo {author} {\bibfnamefont {J.}~\bibnamefont {Conrad}}, \bibinfo {author} {\bibfnamefont {D.}~\bibnamefont {Koser}}, \bibinfo {author} {\bibfnamefont {B.}~\bibnamefont {Koubek}}, \bibinfo {author} {\bibfnamefont {H.}~\bibnamefont {Podlech}}, \bibinfo {author} {\bibfnamefont {U.}~\bibnamefont {Ratzinger}}, \bibinfo {author} {\bibfnamefont {M.}~\bibnamefont {Schuett}}, \bibinfo {author} {\bibfnamefont {J.}~\bibnamefont {Smolsky}}, \bibinfo {author} {\bibfnamefont {M.}~\bibnamefont {Syha}}, \bibinfo {author} {\bibfnamefont {L.}~\bibnamefont {Waites}}, \ and\ \bibinfo {author} {\bibfnamefont {D.}~\bibnamefont {Winklehner}},\ }in\ \href {\doibase 10.18429/JACOW-IPAC2021-THPAB167} {\emph {\bibinfo {booktitle} {Proceedings of the 12th {{International Particle Accelerator Conference}}}}},\ Vol.\ \bibinfo {volume} {IPAC2021}\ (\bibinfo  {publisher} {JACoW Publishing, Geneva, Switzerland},\ \bibinfo {year} {2021})\ pp.\
  \bibinfo {pages} {3 pages, 1.017 MB}\BibitemShut {NoStop}%
\bibitem [{\citenamefont {Stetson}\ \emph {et~al.}(1992)\citenamefont {Stetson}, \citenamefont {Adam}, \citenamefont {Humbel}, \citenamefont {Joho},\ and\ \citenamefont {Stammbach}}]{stetson:vortex}%
  \BibitemOpen
  \bibfield  {author} {\bibinfo {author} {\bibfnamefont {J.}~\bibnamefont {Stetson}}, \bibinfo {author} {\bibfnamefont {S.}~\bibnamefont {Adam}}, \bibinfo {author} {\bibfnamefont {M.}~\bibnamefont {Humbel}}, \bibinfo {author} {\bibfnamefont {W.}~\bibnamefont {Joho}}, \ and\ \bibinfo {author} {\bibfnamefont {T.}~\bibnamefont {Stammbach}},\ }in\ \href@noop {} {\emph {\bibinfo {booktitle} {13th International Conference on Cyclotrons and their Applications}}}\ (\bibinfo {year} {1992})\ p.~\bibinfo {pages} {4}\BibitemShut {NoStop}%
\bibitem [{\citenamefont {Baumgarten}(2011)}]{baumgartenTransverselongitudinalCouplingSpace2011}%
  \BibitemOpen
  \bibfield  {author} {\bibinfo {author} {\bibfnamefont {C.}~\bibnamefont {Baumgarten}},\ }\href {\doibase 10.1103/PhysRevSTAB.14.114201} {\bibfield  {journal} {\bibinfo  {journal} {Phys. Rev. ST Accel. Beams}\ }\textbf {\bibinfo {volume} {14}},\ \bibinfo {pages} {114201} (\bibinfo {year} {2011})}\BibitemShut {NoStop}%
\bibitem [{\citenamefont {Winklehner}\ \emph {et~al.}(2022{\natexlab{a}})\citenamefont {Winklehner}, \citenamefont {Conrad}, \citenamefont {Schoen}, \citenamefont {Yampolskaya}, \citenamefont {Adelmann}, \citenamefont {Mayani},\ and\ \citenamefont {Muralikrishnan}}]{winklehnerOrdermagnitudeBeamCurrent2022}%
  \BibitemOpen
  \bibfield  {author} {\bibinfo {author} {\bibfnamefont {D.}~\bibnamefont {Winklehner}}, \bibinfo {author} {\bibfnamefont {J.~M.}\ \bibnamefont {Conrad}}, \bibinfo {author} {\bibfnamefont {D.}~\bibnamefont {Schoen}}, \bibinfo {author} {\bibfnamefont {M.}~\bibnamefont {Yampolskaya}}, \bibinfo {author} {\bibfnamefont {A.}~\bibnamefont {Adelmann}}, \bibinfo {author} {\bibfnamefont {S.}~\bibnamefont {Mayani}}, \ and\ \bibinfo {author} {\bibfnamefont {S.}~\bibnamefont {Muralikrishnan}},\ }\href {\doibase 10.1088/1367-2630/ac5001} {\bibfield  {journal} {\bibinfo  {journal} {New Journal of Physics}\ }\textbf {\bibinfo {volume} {24}},\ \bibinfo {pages} {023038} (\bibinfo {year} {2022}{\natexlab{a}})}\BibitemShut {NoStop}%
\bibitem [{\citenamefont {Snead}\ \emph {et~al.}(2023)\citenamefont {Snead}, \citenamefont {Winklehner}, \citenamefont {Whyte}, \citenamefont {Zinkle}, \citenamefont {Hartwig},\ and\ \citenamefont {Sprouster}}]{sneadEnablingMultiPurposeHighEnergy2023}%
  \BibitemOpen
  \bibfield  {author} {\bibinfo {author} {\bibfnamefont {L.~L.}\ \bibnamefont {Snead}}, \bibinfo {author} {\bibfnamefont {D.}~\bibnamefont {Winklehner}}, \bibinfo {author} {\bibfnamefont {D.}~\bibnamefont {Whyte}}, \bibinfo {author} {\bibfnamefont {S.}~\bibnamefont {Zinkle}}, \bibinfo {author} {\bibfnamefont {Z.}~\bibnamefont {Hartwig}}, \ and\ \bibinfo {author} {\bibfnamefont {D.}~\bibnamefont {Sprouster}},\ }\href {\doibase 10.48550/arXiv.2302.09011} {\enquote {\bibinfo {title} {Enabling a {{Multi-Purpose High-Energy Neutron Source Based}} on {{High-Current Compact Cyclotrons}}},}\ } (\bibinfo {year} {2023}),\ \Eprint {http://arxiv.org/abs/2302.09011} {arXiv:2302.09011 [physics]} \BibitemShut {NoStop}%
\bibitem [{\citenamefont {Winklehner}\ \emph {et~al.}(2024{\natexlab{b}})\citenamefont {Winklehner}, \citenamefont {Alonso},\ and\ \citenamefont {Conrad}}]{winklehnerNewFamilyHighcurrent2024}%
  \BibitemOpen
  \bibfield  {author} {\bibinfo {author} {\bibfnamefont {D.}~\bibnamefont {Winklehner}}, \bibinfo {author} {\bibfnamefont {J.~R.}\ \bibnamefont {Alonso}}, \ and\ \bibinfo {author} {\bibfnamefont {J.}~\bibnamefont {Conrad}},\ }\href {\doibase 10.1007/s10967-024-09533-3} {\bibfield  {journal} {\bibinfo  {journal} {Journal of Radioanalytical and Nuclear Chemistry}\ } (\bibinfo {year} {2024}{\natexlab{b}}),\ 10.1007/s10967-024-09533-3}\BibitemShut {NoStop}%
\bibitem [{\citenamefont {Alonso}\ \emph {et~al.}(2019)\citenamefont {Alonso}, \citenamefont {Barlow}, \citenamefont {Conrad},\ and\ \citenamefont {Waites}}]{alonsoMedicalIsotopeProduction2019}%
  \BibitemOpen
  \bibfield  {author} {\bibinfo {author} {\bibfnamefont {J.~R.}\ \bibnamefont {Alonso}}, \bibinfo {author} {\bibfnamefont {R.}~\bibnamefont {Barlow}}, \bibinfo {author} {\bibfnamefont {J.~M.}\ \bibnamefont {Conrad}}, \ and\ \bibinfo {author} {\bibfnamefont {L.~H.}\ \bibnamefont {Waites}},\ }\href {\doibase 10.1038/s42254-019-0095-6} {\bibfield  {journal} {\bibinfo  {journal} {Nature Reviews Physics}\ }\textbf {\bibinfo {volume} {1}},\ \bibinfo {pages} {533} (\bibinfo {year} {2019})}\BibitemShut {NoStop}%
\bibitem [{\citenamefont {Hamid Ait~Abderrahim}(2021)}]{ADS_Transmutation}%
  \BibitemOpen
  \bibfield  {author} {\bibinfo {author} {\bibfnamefont {M.~G.}\ \bibnamefont {Hamid Ait~Abderrahim}},\ }\href {\doibase https://doi.org/10.3390/su132212643} {\bibfield  {journal} {\bibinfo  {journal} {Sustainability}\ }\textbf {\bibinfo {volume} {13}},\ \bibinfo {pages} {12643} (\bibinfo {year} {2021})}\BibitemShut {NoStop}%
\bibitem [{\citenamefont {Winklehner}\ \emph {et~al.}(2017)\citenamefont {Winklehner}, \citenamefont {Adelmann}, \citenamefont {Gsell}, \citenamefont {Kaman},\ and\ \citenamefont {Campo}}]{winklehnerRealisticSimulationsCyclotron2017}%
  \BibitemOpen
  \bibfield  {author} {\bibinfo {author} {\bibfnamefont {D.}~\bibnamefont {Winklehner}}, \bibinfo {author} {\bibfnamefont {A.}~\bibnamefont {Adelmann}}, \bibinfo {author} {\bibfnamefont {A.}~\bibnamefont {Gsell}}, \bibinfo {author} {\bibfnamefont {T.}~\bibnamefont {Kaman}}, \ and\ \bibinfo {author} {\bibfnamefont {D.}~\bibnamefont {Campo}},\ }\href {\doibase 10.1103/PhysRevAccelBeams.20.124201} {\bibfield  {journal} {\bibinfo  {journal} {Physical Review Accelerators and Beams}\ }\textbf {\bibinfo {volume} {20}},\ \bibinfo {pages} {124201} (\bibinfo {year} {2017})}\BibitemShut {NoStop}%
\bibitem [{\citenamefont {Peng}\ \emph {et~al.}(2024)\citenamefont {Peng}, \citenamefont {Ma}, \citenamefont {Cui}, \citenamefont {Wu}, \citenamefont {Jiang}, \citenamefont {Guo},\ and\ \citenamefont {Chen}}]{Peking_IonSource}%
  \BibitemOpen
  \bibfield  {author} {\bibinfo {author} {\bibfnamefont {S.}~\bibnamefont {Peng}}, \bibinfo {author} {\bibfnamefont {T.}~\bibnamefont {Ma}}, \bibinfo {author} {\bibfnamefont {B.}~\bibnamefont {Cui}}, \bibinfo {author} {\bibfnamefont {W.}~\bibnamefont {Wu}}, \bibinfo {author} {\bibfnamefont {Y.}~\bibnamefont {Jiang}}, \bibinfo {author} {\bibfnamefont {Z.}~\bibnamefont {Guo}}, \ and\ \bibinfo {author} {\bibfnamefont {J.}~\bibnamefont {Chen}},\ }\href {\doibase 10.1088/1742-6596/2743/1/012056} {\bibfield  {journal} {\bibinfo  {journal} {Journal of Physics: Conference Series}\ }\textbf {\bibinfo {volume} {2743}},\ \bibinfo {pages} {012056} (\bibinfo {year} {2024})}\BibitemShut {NoStop}%
\bibitem [{\citenamefont {Castro}\ \emph {et~al.}(2016)\citenamefont {Castro}, \citenamefont {Torrisi}, \citenamefont {Celona}, \citenamefont {Mascali}, \citenamefont {Neri}, \citenamefont {Sorbello}, \citenamefont {Leonardi}, \citenamefont {Patti}, \citenamefont {Castorina},\ and\ \citenamefont {Gammino}}]{castroNewH2Source2016}%
  \BibitemOpen
  \bibfield  {author} {\bibinfo {author} {\bibfnamefont {G.}~\bibnamefont {Castro}}, \bibinfo {author} {\bibfnamefont {G.}~\bibnamefont {Torrisi}}, \bibinfo {author} {\bibfnamefont {L.}~\bibnamefont {Celona}}, \bibinfo {author} {\bibfnamefont {D.}~\bibnamefont {Mascali}}, \bibinfo {author} {\bibfnamefont {L.}~\bibnamefont {Neri}}, \bibinfo {author} {\bibfnamefont {G.}~\bibnamefont {Sorbello}}, \bibinfo {author} {\bibfnamefont {O.}~\bibnamefont {Leonardi}}, \bibinfo {author} {\bibfnamefont {G.}~\bibnamefont {Patti}}, \bibinfo {author} {\bibfnamefont {G.}~\bibnamefont {Castorina}}, \ and\ \bibinfo {author} {\bibfnamefont {S.}~\bibnamefont {Gammino}},\ }\href {\doibase 10.1063/1.4960564} {\bibfield  {journal} {\bibinfo  {journal} {Review of Scientific Instruments}\ }\textbf {\bibinfo {volume} {87}},\ \bibinfo {pages} {083303} (\bibinfo {year} {2016})}\BibitemShut {NoStop}%
\bibitem [{\citenamefont {Alonso}\ \emph {et~al.}(2014)\citenamefont {Alonso}, \citenamefont {Calabretta}, \citenamefont {Campo}, \citenamefont {Celona}, \citenamefont {Conrad}, \citenamefont {Martinez}, \citenamefont {Johnson}, \citenamefont {Labrecque}, \citenamefont {Toups}, \citenamefont {Winklehner},\ and\ \citenamefont {Winslow}}]{alonsoCharacterizationCataniaVIS2014}%
  \BibitemOpen
  \bibfield  {author} {\bibinfo {author} {\bibfnamefont {J.~R.}\ \bibnamefont {Alonso}}, \bibinfo {author} {\bibfnamefont {L.}~\bibnamefont {Calabretta}}, \bibinfo {author} {\bibfnamefont {D.}~\bibnamefont {Campo}}, \bibinfo {author} {\bibfnamefont {L.}~\bibnamefont {Celona}}, \bibinfo {author} {\bibfnamefont {J.}~\bibnamefont {Conrad}}, \bibinfo {author} {\bibfnamefont {R.~G.}\ \bibnamefont {Martinez}}, \bibinfo {author} {\bibfnamefont {R.}~\bibnamefont {Johnson}}, \bibinfo {author} {\bibfnamefont {F.}~\bibnamefont {Labrecque}}, \bibinfo {author} {\bibfnamefont {M.~H.}\ \bibnamefont {Toups}}, \bibinfo {author} {\bibfnamefont {D.}~\bibnamefont {Winklehner}}, \ and\ \bibinfo {author} {\bibfnamefont {L.}~\bibnamefont {Winslow}},\ }\href {\doibase 10.1063/1.4850736} {\bibfield  {journal} {\bibinfo  {journal} {Review of Scientific Instruments}\ }\textbf {\bibinfo {volume} {85}},\ \bibinfo {pages} {02A742} (\bibinfo {year} {2014})}\BibitemShut {NoStop}%
\bibitem [{\citenamefont {Alonso}\ \emph {et~al.}(2015)\citenamefont {Alonso}, \citenamefont {Axani}, \citenamefont {Calabretta}, \citenamefont {Campo}, \citenamefont {Celona}, \citenamefont {Conrad}, \citenamefont {Day}, \citenamefont {Castro}, \citenamefont {Labrecque},\ and\ \citenamefont {{D. Winklehner}}}]{alonsoIsoDARHighIntensity2015}%
  \BibitemOpen
  \bibfield  {author} {\bibinfo {author} {\bibfnamefont {J.}~\bibnamefont {Alonso}}, \bibinfo {author} {\bibfnamefont {S.}~\bibnamefont {Axani}}, \bibinfo {author} {\bibfnamefont {L.}~\bibnamefont {Calabretta}}, \bibinfo {author} {\bibfnamefont {D.}~\bibnamefont {Campo}}, \bibinfo {author} {\bibfnamefont {L.}~\bibnamefont {Celona}}, \bibinfo {author} {\bibfnamefont {J.~M.}\ \bibnamefont {Conrad}}, \bibinfo {author} {\bibfnamefont {A.}~\bibnamefont {Day}}, \bibinfo {author} {\bibfnamefont {G.}~\bibnamefont {Castro}}, \bibinfo {author} {\bibfnamefont {F.}~\bibnamefont {Labrecque}}, \ and\ \bibinfo {author} {\bibnamefont {{D. Winklehner}}},\ }\href {\doibase 10.1088/1748-0221/10/10/T10003} {\bibfield  {journal} {\bibinfo  {journal} {Journal of Instrumentation}\ }\textbf {\bibinfo {volume} {10}},\ \bibinfo {pages} {T10003} (\bibinfo {year} {2015})}\BibitemShut {NoStop}%
\bibitem [{\citenamefont {Wu}\ \emph {et~al.}(2019)\citenamefont {Wu}, \citenamefont {Peng}, \citenamefont {Ma}, \citenamefont {Ren}, \citenamefont {Zhang}, \citenamefont {Zhang}, \citenamefont {Jiang}, \citenamefont {Li}, \citenamefont {Xu}, \citenamefont {Zhang}, \citenamefont {Wen}, \citenamefont {Guo},\ and\ \citenamefont {Chen}}]{wu:hydrogen}%
  \BibitemOpen
  \bibfield  {author} {\bibinfo {author} {\bibfnamefont {W.}~\bibnamefont {Wu}}, \bibinfo {author} {\bibfnamefont {S.}~\bibnamefont {Peng}}, \bibinfo {author} {\bibfnamefont {T.}~\bibnamefont {Ma}}, \bibinfo {author} {\bibfnamefont {H.}~\bibnamefont {Ren}}, \bibinfo {author} {\bibfnamefont {J.}~\bibnamefont {Zhang}}, \bibinfo {author} {\bibfnamefont {T.}~\bibnamefont {Zhang}}, \bibinfo {author} {\bibfnamefont {Y.}~\bibnamefont {Jiang}}, \bibinfo {author} {\bibfnamefont {K.}~\bibnamefont {Li}}, \bibinfo {author} {\bibfnamefont {Y.}~\bibnamefont {Xu}}, \bibinfo {author} {\bibfnamefont {A.}~\bibnamefont {Zhang}}, \bibinfo {author} {\bibfnamefont {J.}~\bibnamefont {Wen}}, \bibinfo {author} {\bibfnamefont {Z.}~\bibnamefont {Guo}}, \ and\ \bibinfo {author} {\bibfnamefont {J.}~\bibnamefont {Chen}},\ }\href {\doibase 10.1063/1.5109240} {\bibfield  {journal} {\bibinfo  {journal} {Review of Scientific Instruments}\ }\textbf {\bibinfo {volume} {90}},\ \bibinfo {pages} {101501} (\bibinfo {year} {2019})},\ \bibinfo {note}
  {publisher: American Institute of Physics}\BibitemShut {NoStop}%
\bibitem [{\citenamefont {Schweizer}\ \emph {et~al.}(2014)\citenamefont {Schweizer}, \citenamefont {Ratzinger}, \citenamefont {Klump},\ and\ \citenamefont {Volk}}]{schweizerHighIntensity2002014}%
  \BibitemOpen
  \bibfield  {author} {\bibinfo {author} {\bibfnamefont {W.}~\bibnamefont {Schweizer}}, \bibinfo {author} {\bibfnamefont {U.}~\bibnamefont {Ratzinger}}, \bibinfo {author} {\bibfnamefont {B.}~\bibnamefont {Klump}}, \ and\ \bibinfo {author} {\bibfnamefont {K.}~\bibnamefont {Volk}},\ }\href {\doibase 10.1063/1.4842335} {\bibfield  {journal} {\bibinfo  {journal} {Review of Scientific Instruments}\ }\textbf {\bibinfo {volume} {85}},\ \bibinfo {pages} {02A743} (\bibinfo {year} {2014})}\BibitemShut {NoStop}%
\bibitem [{\citenamefont {Joshi}\ \emph {et~al.}(2009)\citenamefont {Joshi}, \citenamefont {Droba}, \citenamefont {Meusel},\ and\ \citenamefont {Ratzinger}}]{joshi:franz1}%
  \BibitemOpen
  \bibfield  {author} {\bibinfo {author} {\bibfnamefont {N.}~\bibnamefont {Joshi}}, \bibinfo {author} {\bibfnamefont {M.}~\bibnamefont {Droba}}, \bibinfo {author} {\bibfnamefont {O.}~\bibnamefont {Meusel}}, \ and\ \bibinfo {author} {\bibfnamefont {U.}~\bibnamefont {Ratzinger}},\ }\href {\doibase 10.1016/j.nima.2009.05.008} {\bibfield  {journal} {\bibinfo  {journal} {Nuclear Instruments and Methods in Physics Research Section A: Accelerators, Spectrometers, Detectors and Associated Equipment}\ }\textbf {\bibinfo {volume} {606}},\ \bibinfo {pages} {310} (\bibinfo {year} {2009})}\BibitemShut {NoStop}%
\bibitem [{\citenamefont {Ehlers}\ and\ \citenamefont {Leung}(1983)}]{ehlers:multicusp1}%
  \BibitemOpen
  \bibfield  {author} {\bibinfo {author} {\bibfnamefont {K.}~\bibnamefont {Ehlers}}\ and\ \bibinfo {author} {\bibfnamefont {K.}~\bibnamefont {Leung}},\ }\href@noop {} {\bibfield  {journal} {\bibinfo  {journal} {Review of Scientific Instruments}\ }\textbf {\bibinfo {volume} {54}},\ \bibinfo {pages} {677} (\bibinfo {year} {1983})}\BibitemShut {NoStop}%
\bibitem [{\citenamefont {Wutte}\ \emph {et~al.}(1998)\citenamefont {Wutte}, \citenamefont {Freedman}, \citenamefont {Gough}, \citenamefont {Lee}, \citenamefont {Leitner}, \citenamefont {Leung}, \citenamefont {Lyneis}, \citenamefont {Pickard}, \citenamefont {Williams},\ and\ \citenamefont {Xie}}]{wutte:multicusp}%
  \BibitemOpen
  \bibfield  {author} {\bibinfo {author} {\bibfnamefont {D.}~\bibnamefont {Wutte}}, \bibinfo {author} {\bibfnamefont {S.}~\bibnamefont {Freedman}}, \bibinfo {author} {\bibfnamefont {R.}~\bibnamefont {Gough}}, \bibinfo {author} {\bibfnamefont {Y.}~\bibnamefont {Lee}}, \bibinfo {author} {\bibfnamefont {M.}~\bibnamefont {Leitner}}, \bibinfo {author} {\bibfnamefont {K.}~\bibnamefont {Leung}}, \bibinfo {author} {\bibfnamefont {C.}~\bibnamefont {Lyneis}}, \bibinfo {author} {\bibfnamefont {D.}~\bibnamefont {Pickard}}, \bibinfo {author} {\bibfnamefont {M.}~\bibnamefont {Williams}}, \ and\ \bibinfo {author} {\bibfnamefont {Z.}~\bibnamefont {Xie}},\ }\href@noop {} {\bibfield  {journal} {\bibinfo  {journal} {Nuclear Instruments and Methods in Physics Research Section B: Beam Interactions with Materials and Atoms}\ }\textbf {\bibinfo {volume} {142}},\ \bibinfo {pages} {409} (\bibinfo {year} {1998})}\BibitemShut {NoStop}%
\bibitem [{\citenamefont {Lee}\ \emph {et~al.}(1996)\citenamefont {Lee}, \citenamefont {Gough}, \citenamefont {Kunkel}, \citenamefont {Leung}, \citenamefont {Perkins}, \citenamefont {Pickard}, \citenamefont {Sun}, \citenamefont {Vujic}, \citenamefont {Williams},\ and\ \citenamefont {Wutte}}]{lee_multicusp}%
  \BibitemOpen
  \bibfield  {author} {\bibinfo {author} {\bibfnamefont {Y.}~\bibnamefont {Lee}}, \bibinfo {author} {\bibfnamefont {R.}~\bibnamefont {Gough}}, \bibinfo {author} {\bibfnamefont {W.}~\bibnamefont {Kunkel}}, \bibinfo {author} {\bibfnamefont {K.}~\bibnamefont {Leung}}, \bibinfo {author} {\bibfnamefont {L.}~\bibnamefont {Perkins}}, \bibinfo {author} {\bibfnamefont {D.}~\bibnamefont {Pickard}}, \bibinfo {author} {\bibfnamefont {L.}~\bibnamefont {Sun}}, \bibinfo {author} {\bibfnamefont {J.}~\bibnamefont {Vujic}}, \bibinfo {author} {\bibfnamefont {M.}~\bibnamefont {Williams}}, \ and\ \bibinfo {author} {\bibfnamefont {D.}~\bibnamefont {Wutte}},\ }\href@noop {} {\bibfield  {journal} {\bibinfo  {journal} {Nuclear Instruments and Methods in Physics Research Section B: Beam Interactions with Materials and Atoms}\ }\textbf {\bibinfo {volume} {119}},\ \bibinfo {pages} {543} (\bibinfo {year} {1996})}\BibitemShut {NoStop}%
\bibitem [{\citenamefont {Winklehner}\ \emph {et~al.}(2018)\citenamefont {Winklehner}, \citenamefont {Axani}, \citenamefont {Bedard}, \citenamefont {Conrad}, \citenamefont {Corona}, \citenamefont {Hartwell}, \citenamefont {Smolsky}, \citenamefont {Tripathee}, \citenamefont {Waites}, \citenamefont {Weigel}, \citenamefont {Wester},\ and\ \citenamefont {Yampolskaya}}]{winklehnerFirstCommissioningResults2018}%
  \BibitemOpen
  \bibfield  {author} {\bibinfo {author} {\bibfnamefont {D.}~\bibnamefont {Winklehner}}, \bibinfo {author} {\bibfnamefont {S.}~\bibnamefont {Axani}}, \bibinfo {author} {\bibfnamefont {P.}~\bibnamefont {Bedard}}, \bibinfo {author} {\bibfnamefont {J.}~\bibnamefont {Conrad}}, \bibinfo {author} {\bibfnamefont {J.}~\bibnamefont {Corona}}, \bibinfo {author} {\bibfnamefont {F.}~\bibnamefont {Hartwell}}, \bibinfo {author} {\bibfnamefont {J.}~\bibnamefont {Smolsky}}, \bibinfo {author} {\bibfnamefont {A.}~\bibnamefont {Tripathee}}, \bibinfo {author} {\bibfnamefont {L.}~\bibnamefont {Waites}}, \bibinfo {author} {\bibfnamefont {P.}~\bibnamefont {Weigel}}, \bibinfo {author} {\bibfnamefont {T.}~\bibnamefont {Wester}}, \ and\ \bibinfo {author} {\bibfnamefont {M.}~\bibnamefont {Yampolskaya}},\ }in\ \href {\doibase 10.1063/1.5053263} {\emph {\bibinfo {booktitle} {{{AIP Conference Proceedings}}}}},\ Vol.\ \bibinfo {volume} {2011}\ (\bibinfo  {publisher} {American Institute of Physics},\ \bibinfo {year} {2018})\ p.\ \bibinfo
  {pages} {030002}\BibitemShut {NoStop}%
\bibitem [{\citenamefont {Winklehner}\ \emph {et~al.}(2021{\natexlab{b}})\citenamefont {Winklehner}, \citenamefont {Conrad}, \citenamefont {Smolsky},\ and\ \citenamefont {Waites}}]{winklehnerHighcurrentH2Beams2021a}%
  \BibitemOpen
  \bibfield  {author} {\bibinfo {author} {\bibfnamefont {D.}~\bibnamefont {Winklehner}}, \bibinfo {author} {\bibfnamefont {J.~M.}\ \bibnamefont {Conrad}}, \bibinfo {author} {\bibfnamefont {J.}~\bibnamefont {Smolsky}}, \ and\ \bibinfo {author} {\bibfnamefont {L.~H.}\ \bibnamefont {Waites}},\ }\href {\doibase 10.1063/5.0063301} {\bibfield  {journal} {\bibinfo  {journal} {Review of Scientific Instruments}\ }\textbf {\bibinfo {volume} {92}},\ \bibinfo {pages} {123301} (\bibinfo {year} {2021}{\natexlab{b}})}\BibitemShut {NoStop}%
\bibitem [{\citenamefont {Winklehner}\ \emph {et~al.}(2022{\natexlab{b}})\citenamefont {Winklehner}, \citenamefont {Conrad}, \citenamefont {Smolsky}, \citenamefont {Waites},\ and\ \citenamefont {Weigel}}]{winklehnerNewCommissioningResults2022}%
  \BibitemOpen
  \bibfield  {author} {\bibinfo {author} {\bibfnamefont {D.}~\bibnamefont {Winklehner}}, \bibinfo {author} {\bibfnamefont {J.}~\bibnamefont {Conrad}}, \bibinfo {author} {\bibfnamefont {J.}~\bibnamefont {Smolsky}}, \bibinfo {author} {\bibfnamefont {L.}~\bibnamefont {Waites}}, \ and\ \bibinfo {author} {\bibfnamefont {P.}~\bibnamefont {Weigel}},\ }\href {\doibase 10.1088/1742-6596/2244/1/012013} {\bibfield  {journal} {\bibinfo  {journal} {Journal of Physics: Conference Series}\ }\textbf {\bibinfo {volume} {2244}},\ \bibinfo {pages} {012013} (\bibinfo {year} {2022}{\natexlab{b}})}\BibitemShut {NoStop}%
\bibitem [{\citenamefont {Winklehner}\ and\ \citenamefont {Leitner}(2015)}]{winklehnerSpaceChargeCompensation2015}%
  \BibitemOpen
  \bibfield  {author} {\bibinfo {author} {\bibfnamefont {D.}~\bibnamefont {Winklehner}}\ and\ \bibinfo {author} {\bibfnamefont {D.}~\bibnamefont {Leitner}},\ }\href {\doibase 10.1088/1748-0221/10/10/T10006} {\bibfield  {journal} {\bibinfo  {journal} {Journal of Instrumentation}\ }\textbf {\bibinfo {volume} {10}},\ \bibinfo {pages} {T10006} (\bibinfo {year} {2015})}\BibitemShut {NoStop}%
\bibitem [{\citenamefont {Weigel}\ \emph {et~al.}(2023)\citenamefont {Weigel}, \citenamefont {Busza}, \citenamefont {Namazov}, \citenamefont {Park}, \citenamefont {Villarreal}, \citenamefont {Waites},\ and\ \citenamefont {Winklehner}}]{WEIGEL2023168590}%
  \BibitemOpen
  \bibfield  {author} {\bibinfo {author} {\bibfnamefont {P.}~\bibnamefont {Weigel}}, \bibinfo {author} {\bibfnamefont {M.}~\bibnamefont {Busza}}, \bibinfo {author} {\bibfnamefont {A.}~\bibnamefont {Namazov}}, \bibinfo {author} {\bibfnamefont {J.}~\bibnamefont {Park}}, \bibinfo {author} {\bibfnamefont {J.}~\bibnamefont {Villarreal}}, \bibinfo {author} {\bibfnamefont {L.~H.}\ \bibnamefont {Waites}}, \ and\ \bibinfo {author} {\bibfnamefont {D.}~\bibnamefont {Winklehner}},\ }\href {\doibase https://doi.org/10.1016/j.nima.2023.168590} {\bibfield  {journal} {\bibinfo  {journal} {Nuclear Instruments and Methods in Physics Research Section A: Accelerators, Spectrometers, Detectors and Associated Equipment}\ }\textbf {\bibinfo {volume} {1056}},\ \bibinfo {pages} {168590} (\bibinfo {year} {2023})}\BibitemShut {NoStop}%
\bibitem [{\citenamefont {Duckitt}\ and\ \citenamefont {Abraham}(2020)}]{duckitt:cyclotrons2019-tha03}%
  \BibitemOpen
  \bibfield  {author} {\bibinfo {author} {\bibfnamefont {W.}~\bibnamefont {Duckitt}}\ and\ \bibinfo {author} {\bibfnamefont {J.}~\bibnamefont {Abraham}},\ }in\ \href {\doibase 10.18429/JACoW-Cyclotrons2019-THA03} {\emph {\bibinfo {booktitle} {Proc. Cyclotrons'19}}},\ \bibinfo {series and number} {\bibinfo {series} {International Conference on Cyclotrons and their Applications}\ No.~\bibinfo {number} {22}}\ (\bibinfo  {publisher} {JACoW Publishing, Geneva, Switzerland},\ \bibinfo {year} {2020})\ pp.\ \bibinfo {pages} {285--288},\ \bibinfo {note} {https://doi.org/10.18429/JACoW-Cyclotrons2019-THA03}\BibitemShut {NoStop}%
\bibitem [{\citenamefont {Duckitt}\ \emph {et~al.}(2024)\citenamefont {Duckitt}, \citenamefont {Abraham}, \citenamefont {Marcato},\ and\ \citenamefont {Savarese}}]{duckitt:icalepcs2023-fr2bco01}%
  \BibitemOpen
  \bibfield  {author} {\bibinfo {author} {\bibfnamefont {W.}~\bibnamefont {Duckitt}}, \bibinfo {author} {\bibfnamefont {J.}~\bibnamefont {Abraham}}, \bibinfo {author} {\bibfnamefont {D.}~\bibnamefont {Marcato}}, \ and\ \bibinfo {author} {\bibfnamefont {G.}~\bibnamefont {Savarese}},\ }in\ \href {\doibase 10.18429/JACoW-ICALEPCS2023-FR2BCO01} {\emph {\bibinfo {booktitle} {Proc. ICALEPCS'23}}},\ \bibinfo {series and number} {\bibinfo {series} {International Conference on Accelerator and Large Experimental Physics Control Systems}\ No.~\bibinfo {number} {19}}\ (\bibinfo  {publisher} {JACoW Publishing, Geneva, Switzerland},\ \bibinfo {year} {2024})\ pp.\ \bibinfo {pages} {1643--1649}\BibitemShut {NoStop}%
\bibitem [{EPI()}]{EPICS}%
  \BibitemOpen
  \href@noop {} {\enquote {\bibinfo {title} {Epics - experimental physics and industrial control system},}\ }\bibinfo {howpublished} {\url{https://epics-controls.org/}}\BibitemShut {NoStop}%
\bibitem [{mis()}]{mist1-control-system-github}%
  \BibitemOpen
  \href@noop {} {\enquote {\bibinfo {title} {\texttt{rfq-dip-epics}},}\ }\bibinfo {howpublished} {\url{https://github.com/DanielWinklehner/rfq-dip-epics}}\BibitemShut {NoStop}%
\bibitem [{\citenamefont {Kalvas}\ \emph {et~al.}(2010)\citenamefont {Kalvas}, \citenamefont {Tarvainen}, \citenamefont {Ropponen}, \citenamefont {Steczkiewicz}, \citenamefont {\"Arje},\ and\ \citenamefont {Clark}}]{kalvas:ibsimu}%
  \BibitemOpen
  \bibfield  {author} {\bibinfo {author} {\bibfnamefont {T.}~\bibnamefont {Kalvas}}, \bibinfo {author} {\bibfnamefont {O.}~\bibnamefont {Tarvainen}}, \bibinfo {author} {\bibfnamefont {T.}~\bibnamefont {Ropponen}}, \bibinfo {author} {\bibfnamefont {O.}~\bibnamefont {Steczkiewicz}}, \bibinfo {author} {\bibfnamefont {J.}~\bibnamefont {\"Arje}}, \ and\ \bibinfo {author} {\bibfnamefont {H.}~\bibnamefont {Clark}},\ }\href {\doibase 10.1063/1.3258608} {\bibfield  {journal} {\bibinfo  {journal} {Review of Scientific Instruments}\ }\textbf {\bibinfo {volume} {81}},\ \bibinfo {pages} {02B703} (\bibinfo {year} {2010})},\ \bibinfo {note} {publisher: American Institute of Physics}\BibitemShut {NoStop}%
\bibitem [{\citenamefont {Friedman}\ \emph {et~al.}(2014)\citenamefont {Friedman}, \citenamefont {Cohen}, \citenamefont {Grote}, \citenamefont {Lund}, \citenamefont {Sharp}, \citenamefont {Vay}, \citenamefont {Haber},\ and\ \citenamefont {Kishek}}]{friedman:warp}%
  \BibitemOpen
  \bibfield  {author} {\bibinfo {author} {\bibfnamefont {A.}~\bibnamefont {Friedman}}, \bibinfo {author} {\bibfnamefont {R.~H.}\ \bibnamefont {Cohen}}, \bibinfo {author} {\bibfnamefont {D.~P.}\ \bibnamefont {Grote}}, \bibinfo {author} {\bibfnamefont {S.~M.}\ \bibnamefont {Lund}}, \bibinfo {author} {\bibfnamefont {W.~M.}\ \bibnamefont {Sharp}}, \bibinfo {author} {\bibfnamefont {J.-L.}\ \bibnamefont {Vay}}, \bibinfo {author} {\bibfnamefont {I.}~\bibnamefont {Haber}}, \ and\ \bibinfo {author} {\bibfnamefont {R.~A.}\ \bibnamefont {Kishek}},\ }\href {\doibase 10.1109/TPS.2014.2308546} {\bibfield  {journal} {\bibinfo  {journal} {IEEE Transactions on Plasma Science}\ }\textbf {\bibinfo {volume} {42}},\ \bibinfo {pages} {1321} (\bibinfo {year} {2014})},\ \bibinfo {note} {conference Name: IEEE Transactions on Plasma Science}\BibitemShut {NoStop}%
\bibitem [{\citenamefont {Koser}\ \emph {et~al.}(2022)\citenamefont {Koser}, \citenamefont {Waites}, \citenamefont {Winklehner}, \citenamefont {Frey}, \citenamefont {Adelmann},\ and\ \citenamefont {Conrad}}]{koserInputBeamMatching2022}%
  \BibitemOpen
  \bibfield  {author} {\bibinfo {author} {\bibfnamefont {D.}~\bibnamefont {Koser}}, \bibinfo {author} {\bibfnamefont {L.}~\bibnamefont {Waites}}, \bibinfo {author} {\bibfnamefont {D.}~\bibnamefont {Winklehner}}, \bibinfo {author} {\bibfnamefont {M.}~\bibnamefont {Frey}}, \bibinfo {author} {\bibfnamefont {A.}~\bibnamefont {Adelmann}}, \ and\ \bibinfo {author} {\bibfnamefont {J.}~\bibnamefont {Conrad}},\ }\href@noop {} {\bibfield  {journal} {\bibinfo  {journal} {Frontiers in Physics}\ }\textbf {\bibinfo {volume} {10}} (\bibinfo {year} {2022})}\BibitemShut {NoStop}%
\bibitem [{\citenamefont {Villarreal}\ \emph {et~al.}(2024)\citenamefont {Villarreal}, \citenamefont {Winklehner}, \citenamefont {Koser},\ and\ \citenamefont {Conrad}}]{villarrealNeuralNetworksEffective2024}%
  \BibitemOpen
  \bibfield  {author} {\bibinfo {author} {\bibfnamefont {J.}~\bibnamefont {Villarreal}}, \bibinfo {author} {\bibfnamefont {D.}~\bibnamefont {Winklehner}}, \bibinfo {author} {\bibfnamefont {D.}~\bibnamefont {Koser}}, \ and\ \bibinfo {author} {\bibfnamefont {J.~M.}\ \bibnamefont {Conrad}},\ }\href {\doibase 10.1088/2632-2153/ad3a30} {\bibfield  {journal} {\bibinfo  {journal} {Machine Learning: Science and Technology}\ }\textbf {\bibinfo {volume} {5}},\ \bibinfo {pages} {025009} (\bibinfo {year} {2024})}\BibitemShut {NoStop}%
\bibitem [{\citenamefont {Adelmann}(2019)}]{doi:10.1137/16M1061928}%
  \BibitemOpen
  \bibfield  {author} {\bibinfo {author} {\bibfnamefont {A.}~\bibnamefont {Adelmann}},\ }\href {\doibase 10.1137/16M1061928} {\bibfield  {journal} {\bibinfo  {journal} {SIAM/ASA Journal on Uncertainty Quantification}\ }\textbf {\bibinfo {volume} {7}},\ \bibinfo {pages} {383} (\bibinfo {year} {2019})},\ \Eprint {http://arxiv.org/abs/https://doi.org/10.1137/16M1061928} {https://doi.org/10.1137/16M1061928} \BibitemShut {NoStop}%
\bibitem [{\citenamefont {Kaiser}\ \emph {et~al.}(2024)\citenamefont {Kaiser}, \citenamefont {Xu}, \citenamefont {Eichler}, \citenamefont {Santamaria~Garcia}, \citenamefont {Stein}, \citenamefont {Br{\"u}ndermann}, \citenamefont {Kuropka}, \citenamefont {Dinter}, \citenamefont {Mayet}, \citenamefont {Vinatier}, \citenamefont {Burkart},\ and\ \citenamefont {Schlarb}}]{Kaiser2024-xw}%
  \BibitemOpen
  \bibfield  {author} {\bibinfo {author} {\bibfnamefont {J.}~\bibnamefont {Kaiser}}, \bibinfo {author} {\bibfnamefont {C.}~\bibnamefont {Xu}}, \bibinfo {author} {\bibfnamefont {A.}~\bibnamefont {Eichler}}, \bibinfo {author} {\bibfnamefont {A.}~\bibnamefont {Santamaria~Garcia}}, \bibinfo {author} {\bibfnamefont {O.}~\bibnamefont {Stein}}, \bibinfo {author} {\bibfnamefont {E.}~\bibnamefont {Br{\"u}ndermann}}, \bibinfo {author} {\bibfnamefont {W.}~\bibnamefont {Kuropka}}, \bibinfo {author} {\bibfnamefont {H.}~\bibnamefont {Dinter}}, \bibinfo {author} {\bibfnamefont {F.}~\bibnamefont {Mayet}}, \bibinfo {author} {\bibfnamefont {T.}~\bibnamefont {Vinatier}}, \bibinfo {author} {\bibfnamefont {F.}~\bibnamefont {Burkart}}, \ and\ \bibinfo {author} {\bibfnamefont {H.}~\bibnamefont {Schlarb}},\ }\href@noop {} {\bibfield  {journal} {\bibinfo  {journal} {Sci. Rep.}\ }\textbf {\bibinfo {volume} {14}},\ \bibinfo {pages} {15733} (\bibinfo {year} {2024})}\BibitemShut {NoStop}%
\bibitem [{\citenamefont {Rajput}\ \emph {et~al.}(2025)\citenamefont {Rajput}, \citenamefont {Lin}, \citenamefont {Edelen}, \citenamefont {Blokland},\ and\ \citenamefont {Schram}}]{rajput2025outlookdeployablecontinuallearning}%
  \BibitemOpen
  \bibfield  {author} {\bibinfo {author} {\bibfnamefont {K.}~\bibnamefont {Rajput}}, \bibinfo {author} {\bibfnamefont {S.}~\bibnamefont {Lin}}, \bibinfo {author} {\bibfnamefont {A.}~\bibnamefont {Edelen}}, \bibinfo {author} {\bibfnamefont {W.}~\bibnamefont {Blokland}}, \ and\ \bibinfo {author} {\bibfnamefont {M.}~\bibnamefont {Schram}},\ }\href {https://arxiv.org/abs/2504.03793} {\enquote {\bibinfo {title} {Outlook towards deployable continual learning for particle accelerators},}\ } (\bibinfo {year} {2025}),\ \Eprint {http://arxiv.org/abs/2504.03793} {arXiv:2504.03793 [cs.LG]} \BibitemShut {NoStop}%
\end{thebibliography}%

\end{document}